\documentclass[%
 pra,reprint,
 amsmath,amssymb,
 aps,
floatfix,
]{revtex4-2}

\usepackage{physics}
\usepackage{graphicx}
\usepackage{subcaption}
\usepackage{quantikz}

\usepackage{dcolumn}
\usepackage{bm}

\usepackage{hyperref}
\hypersetup{
    colorlinks=true,       
    linkcolor=cyan,          
    citecolor=magenta,        
    filecolor=magenta,      
    urlcolor=cyan,           
    runcolor=cyan
}

\begin{document}

\preprint{APS/123-QED}

\title{Continuous quantum correction on Markovian and Non-Markovian models}

\author{Juan Garcia Nila }
 \email{jgarcian@usc.edu}

\author{Todd A. Brun}%
 \email{tbrun@usc.edu}
\affiliation{%
 University of Southern California, Viterbi School of Engineering\\
}%

\date{\today}

\begin{abstract}

We investigate continuous quantum error correction, comparing performance under a Markovian error model to two distinct non-Markovian models. The first non-Markovian model involves an interaction Hamiltonian between the system and an environmental qubit via an $X$-$X$ coupling, with a ``cooling'' bath acting on the environment qubit. This model is known to exhibit abrupt transitions between Markovian and non-Markovian behavior \cite{Pang2017}. The second non-Markovian model uses the post-Markovian master equation (PMME), which represents the bath correlation through a memory kernel; we consider an exponentially decaying kernel and both underdamped and overdamped dynamics. We systematically compare these non-Markovian error models against the Markovian case and against each other, for a variety of different codes. We start with a single qubit, which can be solved analytically. We then consider the three-qubit repetition code and the five-qubit ``perfect'' code. In all cases, we find that the fidelity decays more rapidly in the Markovian case than in either non-Markovian model, suggesting that continuous quantum error correction has enhanced performance against non-Markovian noise. We attribute this difference to the presence of a quantum Zeno regime in both non-Markovian models.
\end{abstract}

\maketitle


\section{Introduction}
\label{sec:intro}

Quantum computation offers advantages over classical computation, but it is highly susceptible to information loss due to interactions with the environment, a process known as decoherence \cite{Nielsen}. Errors during computation can significantly affect the results, making Quantum Error Correction (QEC) \cite{BrunQEC,Brun2019,Chatterje2023} and Fault Tolerance (FT) \cite{Gottesman1998} crucial. Despite suppression methods such as quantum control and dynamical decoupling, error probabilities accumulate over time. To mitigate this, Paz and Zurek proposed continuous quantum error correction (CQEC) in 1998 \cite{Zurek1998}, showing by a Markovian master equation that error correction operations could in principle be applied continuously (or in arbitrarily short time intervals compared to other time scales in the system).

In 2002, Ahn, Doherty, and Landahl (ADL) \cite{Ahn2002} described a CQEC protocol that continuously monitors values of a set of stabilizers and applies corresponding feedback. While successful, this approach requires a filtering procedure that is computationally difficult and not optimal. Alternative proposals enhance performance using coarse-grained state estimates \cite{Sarovar2004}, strategies based on classical hybrid control theory \cite{Mabuchi2009}, and controls driven by Brownian motion \cite{Cardona2019}. These techniques have been applied even to the nine-qubit Bacon-Shor code \cite{Atalaya2020}.

Experimental demonstrations of continuous measurements for real-time feedback have been proposed in optical cavities \cite{Raghunathan09,Raghunathan10,Lanka2023} and achieved in superconducting qubit architectures \cite{Minev2019, DeLange2014}. Recently, a demonstration of the bit-flip code using parity measurement in three transmons was realized \cite{Livingston2022}.

It is generally assumed that environmental interaction is memoryless (Markovian) when the system is weakly coupled to a bath, providing a significant simplification to the theoretical description. Alternatives for error correction using an open-quantum-system approach include error suppression, which involves encoding the desired Hamiltonian by adding a penalty term such that any state outside the code space will be at higher energy than the code states \cite{Marvian2017,Sarovar2013}. Another common approach is dynamical decoupling, which implements periodic sequences of pulses to remove unwanted bath coupling \cite{Viola1999,Ezzell2023}. Both of these approaches, however, are ineffective against Markovian noise. Standard QEC can be used against both Markovian and non-Markovian noise. A recent analysis has also applied the Petz recovery map to non-Markovian noise \cite{Shrikant2024}.

Superconducting qubit systems can exhibit non-Markovian effects. Imperfect fabrication in superconducting qubits can lead to two-level-systems noise \cite{Agarwal2024}, crosstalk between qubits \cite{Vinay2022}, or gate operation fidelities conditioned by previous gate sequences \cite{Zhang2021}. This type of noise occurs in many superconducting designs, including fluxonium qubits  \cite{zhuang2025}. An experimental demonstration using a superconducting qubit processor, where they engineered the non-Markovianity of the environment by populating its readout cavity with thermal photons was shown in \cite{Gaikwad2024}. Since many realistic systems have non-Markovian noise, it is useful to understand how different approaches to error correction or suppression perform in that case. This requires more-or-less tractable models of a non-Markovian environment.

Oreshkov and Brun \cite{Oreshkov2007} studied CQEC with a very simple non-Markovian noise model, where each qubit of the code interacted independently with a single environment qubit. Pang et al. studied a more sophisticated version of this model \cite{Pang2017}, with a simple $X$-$X$ system-bath coupling between each code qubit and one environment qubit, where each environment qubit is itself decohered by a Markovian ``cooling bath.'' The competition between these effects leads to a richer behavior, and can show a transition between Markovian and non-Markovian behavior, depending on the relative strength of the two terms.

To address open dynamics where the noise source is unknown, Shabani and Lidar introduced the Post-Markovian Master Equation (PMME) \cite{Shabani2005}, which includes phenomenological bath memory effects and can be analytically solved in some cases via Laplace transforms. Its non-Markovianity has been analyzed with a measure based on non-completely positive dynamics \cite{Sutherland2018}.

This formalism offers an alternative to the Nakajima-Zwanzig equation \cite{Zwanzig1960} or the time-convolutionless Lindblad equation, which require explicit bath information; consider, for example, the spin-boson model \cite{Zihan2024}, which can describe transmon qubit dynamics \cite{Vaaranta2022}. Experimental quantum simulations of the spin-boson model have been done using a superconducting microwave cavity \cite{Leppakangas2018} and using the trapped-ion architecture \cite{Sun2025}.

Other Non-Markovian approximations include the time-convolutionless Lindblad (TCL) equation \cite{DragomirTCL4,DragomirTCL6}, on the parametric approximation \cite{Teretenkov2025}, linear master equation \cite{Dann2025} or even using the  non-Markovian quantum stochastic diferential equation (NMQSD) \cite{Li2025Spin} and using numerical packages like in QuTiP using hierarchical equations of motion (HEOM) \cite{QutipBofin} and the Hamiltonian open quantum system toolkit (HOQST) \cite{HOQSTpaper}.

Oreshkov's work on CQEC with non-Markovian errors \cite{Oreshkov2007} showed ehanced error-correcting fidelity relative to Markovian dynamics, due to the existence of a Zeno regime: the continuous monitoring in CQEC suppresses the transitions that cause errors. Hsu and Brun studied a similar protocol, but reduced the number of ancillas required by an exponential factor \cite{Hsu2016}. Continuous monitoring even in a quantum error-detecting code, like the four-qubit Bacon-Shor code, can suppress non-Markovian, but not Markovian, errors \cite{YiHsiang2020}.

This work aims to generalize previous studies and provide insights into how non-Markovian interactions can be studied in more general stabilizer codes, with more sophisticated error models. Due to the analytical simplifications, we focus on smaller codes, and treat one, three, and five-qubit codes in detail.

The structure of this paper is as follows: In Section~\ref{sec:One_qubit_CQEC}, we describe the error-correcting map for one qubit initially in a pure ground state, subject to bit-flip noise, in the Markovian case. For the non-Markovian models, we use the $X$-$X$ coupling with a cooling bath introduced by \cite{Pang2017} and the PMME with two different memory kernels. In Section~\ref{sec:Fidelity_ZE}, we compare the fidelity of the models described in Section~\ref{sec:One_qubit_CQEC}, presenting figures of their dynamics and short-time expansions. In Section~\ref{sec:More_general_codes}, we generalize the error-correcting procedure and master equations to general $[[n,k]]$ codes (where $n$ physical qubits encode $k$ logical qubits), specifically focusing on the three-qubit repetition code and the five-qubit (perfect) code. A summary of the results is given in Section~\ref{sec:conclusions}. Additional details are provided in the Appendix.

\section{One qubit with continuous quantum error correction}
\label{sec:One_qubit_CQEC}

First we analyze the simplest toy model, where we maintain a single-qubit system in its pure ground state $\ket{0}^S$. 

From the stabilizer formalism, we can construct a continuous error-correcting generator. The procedure consists in measuring the stabilizer $Z$ and then applying a unitary correction accordingly. If the measurement yields $\ket{0}$ we do nothing; while if it yields $\ket{1}$ we apply the $X$ operation. The corresponding completely positive (CP) map is 
\begin{equation}\label{eq:cp_1q_correction}
\Phi(\rho)=\ket{0}\bra{0}\rho \ket{0}\bra{0}+X \ket{1}\bra{1}\rho \ket{1}\bra{1}X.
\end{equation}
We can then define a continuous-time error-correcting generator:
\begin{equation}\label{Gamma1qubit}
\Gamma(\rho^S)=\Phi(\rho^S)-\rho^S . 
\end{equation}
Empirically, this procedure can be done using weak measurements of the stabilizer $Z$ and applying weak rotations around the $X-$axis in the Bloch sphere \cite{BrunQEC}. In the ADL approach \cite{Ahn2002}, these rotations are feedback operations conditioned on a measurement record.

In this section, we derive analytical expressions for continuous correction of the one-qubit code for three different error models: the Markovian case, an $X$-$X$ coupling between the system and a bath qubit with a Markovian cooling process, and the PMME.

\subsection{Markovian bit-flip noise}

The simplest scenario, analogous to the classical bit-flip error model, involves one qubit being perturbed by bit-flip decoherence. To describe the dynamics of this simple model, we use a master equation for the system density matrix $\rho^S$:
\begin{equation}\label{simplest}
\partial_t \rho^S=\gamma \mathcal{D}(X)\rho^S+\eta \Gamma(\rho^S),
\end{equation}
where $\mathcal{D}(X)$ is the Lindblad generator acting over $\rho^S$  describing a Markovian $X$-bit flip process,
\begin{equation}\label{Xlindbaldian}
	\mathcal{D}(X)\rho^S = X \rho^S X -\rho^S ,
\end{equation}
and the system is continuously corrected by the generator $\Gamma(\rho^S)$ defined in Eq.~(\ref{Gamma1qubit}) and associated with the CP map $\Phi(\rho^S)$ in Eq.~(\ref{eq:cp_1q_correction}). The parameters $\gamma$ and $\eta$ represent the decoherence and error-correcting rates, respectively. 

Using the Pauli basis, we assume that the initial system qubit state for the master equation (\ref{simplest}) is 
\begin{equation}\label{init}
	\rho_0^S = \frac{1}{2}(I^S+x_0 X + y_0 Y + z_0 Z) .
\end{equation}
In Appendix~\ref{app:Markov_1q} we derive in detail the solution $\rho_t$ at an arbitrary time $t>0$, which is
\begin{eqnarray}\label{solsimplest}
\rho_t^S &=& \frac{1}{2}\biggl[I + x_0 e^{-\eta t} X
+ e^{-(2\gamma+\eta) t}(y_0 Y + z_0 Z)\nonumber \\ 
&& + \frac{\eta}{2\gamma+\eta} \left(1-e^{-(2\gamma+\eta) t}\right)Z \biggr].
\end{eqnarray}
We can verify the stationary limits $\rho^S_\infty$ at long times. When the dissipation term is greater than the correction, we obtain a completely mixed state:
\[
\lim_{\eta/\gamma\rightarrow 0}
\rho_\infty^S = \frac{1}{2}I^S .
\]
Conversely, when the correction is much greater than the dissipation term, we recover the desired pure state:
\[
\lim_{\eta/\gamma\rightarrow \infty}
\rho_\infty^S = \frac{1}{2}\left(I+Z\right)
= \ket{0}\!\bra{0} .
\]

\subsection{System-bath $X$-$X$ coupling with a cooling bath}

In this section, we describe the model introduced by Pang \cite{Pang2017} of an open one-qubit system interacting with one bath qubit via the $X$-$X$ coupling using the interaction Hamiltonian
\begin{equation}\label{XXcoup}
	H_{int}=\alpha X^S\otimes X^B.
\end{equation}
Although the model appears simplistic for describing real noise, a common source of noise in superconducting architectures is coupling to two-level systems in the environment. While the nature of these systems is not fully understood, they in many cases act like perfect qubits. Another common source of noise is crosstalk, usually $Z$-$Z$ coupling between two qubits. (For instance in \cite{Vinay2022}, the authors suppress the crosstalk using dynamical decoupling.) $Z$-$Z$ crosstalk behaves similarly to the model we consider in this paper if we modify our generator to the $X$-basis by interchanging $\ket{0}$ and $\ket{1}$ with the $\ket{\pm}$ basis.

The dynamics induced by this Hamiltonian are non-Markovian. To simulate Markovian dynamics, Pang proposed a quick restoration of the bath qubit into its ground state, which would represent the effect of a large reservoir at zero temperature. (It could also represent other physical mechanisms, such as amplitude damping by spontaneous emission into the environment.) This model has also been studied for memory strength, which quantifies detectable conditional future-history correlations \cite{Taranto2021}, and a similar cooling approach has been used as a continuous error correcting procedure \cite{Sarovarcool}.

The cooling process on the bath qubit is modeled by a dissipation term $\mathcal{D}[\sigma_-^B]$ at a cooling rate $\kappa$:
\begin{equation}\label{eq:cool_disip}
\mathcal{D}[\sigma_-^B]\rho_t^{SB}=\sigma_-^B\rho_t^{SB}\sigma_+^B-\frac{1}{2}\{\sigma_+^B\sigma_-^B,\rho^{SB}_t\} .
\end{equation}
Here we use the operators $\sigma_{-}=\ket0\bra1$ and $\sigma_{+}=\ket1\bra0 = \sigma_{-}^\dagger$, which transform $\ket{1}\rightarrow \ket{0}$ and vice versa.

We combine this model with the error correction generator from Eq.~ (\ref{Gamma1qubit}) acting on the system to obtain a master equation on the joint system-bath state $\rho_t^{SB}$:
\begin{equation}\label{couplingbathgamma}
	\partial_t \rho_t^{SB} = -i\alpha[X^S \otimes  X^B,\rho_t^{SB}] + \kappa \mathcal{D}[\sigma_-^B] + \eta \Gamma^S(\rho)
\end{equation}
We solve this assuming an initial separable state where the system state is given by Eq.~(\ref{init}), while the bath qubit state is $\ket{0}^B$. The details of the calculation are given in Appendix~\ref{app:XX_1q}. By tracing out the bath qubit on the system-bath solution at time $t$, we obtain the reduced density matrix for the system:
\begin{equation}\label{evolr}
	\rho_t^S=\frac{1}{2}(I^S + x_0 e^{-\eta t} X^S + \mathfrak{C}(t)(y_0 Y^S + z_0 Z^S) + \mathfrak{D}(t) Z^S) ,
\end{equation}
where the coefficients $\mathfrak{C}(t)$ and $\mathfrak{D}(t)$ are given by
\begin{widetext}
\begin{equation}\label{eq:c_coeff_XX}
\mathfrak{C}(t)=\begin{cases}
		e^{-\left(\eta+\kappa/4 \right)t} \left( \frac{\kappa \sin \left((t/4)\sqrt{64\alpha^2-\kappa^2}\right) }{\sqrt{64\alpha^2-\kappa^2}}) +\cos\left(\frac{t}{4} \sqrt{64\alpha^2-\kappa^2}\right)\right) , & \kappa^2<64 \alpha^2 , \\
		e^{-\left(\eta + \kappa/4\right)t} \left(\frac{\kappa \sinh \left((t/4)\sqrt{\kappa^2-64\alpha^2}\right) }{\sqrt{\kappa^2-64\alpha^2}} + \cosh\left(\frac{t}{4} \sqrt{\kappa^2-64\alpha^2}\right)\right) , & \kappa^2 > 64 \alpha^2 , \\
		e^{-\left(\eta+2\alpha\right)t}\left(1+2\alpha t\right) , & \kappa^2 = 64 \alpha^2 ,
	\end{cases}
\end{equation}
and,
\begin{equation}\label{eq:d_coeff_XX}
\mathfrak{D}(t) = \begin{cases}
		\frac{\eta(\kappa+2\eta)} {\eta(\kappa+2\eta)+8\alpha^2} \biggl(1 + e^{-\left(\eta+\kappa/4\right)t} \left(\left(\frac{32\alpha^2}{k+2\eta}-\kappa\right)
	\frac{\sin((t/4)\sqrt{64\alpha^2-\kappa^2})}{\sqrt{64\alpha^2-\kappa^2}} - \cos\left(\frac{t}{4}\sqrt{64\alpha^2-\kappa^2}\right)\right) , & \kappa^2<64 \alpha^2 , \\
	\frac{\eta(\kappa+2\eta)}{\eta(\kappa+2\eta)+8\alpha^2} \biggl(1 + e^{-\left(\eta + \kappa/4\right)t} \left(\left(\frac{32\alpha^2}{k+2\eta} - \kappa\right) \frac{\sinh((t/4) \sqrt{\kappa^2-64\alpha^2})}{\sqrt{\kappa^2-64\alpha^2}} - \cosh\left(\frac{t}{4} \sqrt{\kappa^2-64\alpha^2}\right) \right) , & \kappa^2>64 \alpha^2 , \\
    \frac{\eta(\eta+4\alpha)}{(\eta+2\alpha)^2} \left(1 + 	e^{-\left(\eta+2\alpha\right)t} 
		\left(\left(\frac{4\alpha^2}{4\alpha+\eta} - 2\alpha\right)t - 1 \right)\right) , & \kappa^2=64 \alpha^2.
	\end{cases}
\end{equation}
\end{widetext}
These equations generalize the coefficients in \cite{Pang2017} for the case of no correction ($\eta=0$). In the following subsection, we demonstrate, using a measurement based on the CP-divisibility theorem, that the dynamics of the system qubit is Markovian when $\kappa^2\geq 64 \alpha^2$ and non-Markovian when $\kappa^2< 64 \alpha^2$.

Allowing the system to evolve over a long time leads to a stationary state along the $z-$axis of the Bloch sphere
\begin{equation}\label{asympevolrho}
	\lim_{t\rightarrow \infty}\rho_t^S = \frac{1}{2}\left(I^S + \frac{\eta(\kappa+2\eta)}{\eta(\kappa+2\eta)+8\alpha^2} Z^S\right).
\end{equation}
For strong correction rate $\eta\gg 1$, the asymptotic corrected state $\rho^s\rightarrow (1/2)(I^S + Z^S)=\ket{0}\bra{0}$.

\subsubsection{Trace distance measure}
A measure $\mathcal{N}$ for non-Markovianity, proposed by Breuer \cite{Breuer2012}, is based on the trace distance between two quantum states that evolve by the system dynamics:
\begin{equation}\label{measure}
	\mathcal{N}=\max_{\rho_{1,2}(0)} \int_{\partial_{t'}d_{Tr}>0} \partial_{t'}d_{Tr}(\rho_1(t'),\rho_2(t'))dt'
\end{equation}
where $d_{Tr}(\rho_1(t),\rho_2(t))$ is the trace distance:
\begin{equation}\label{traced}
	d_{Tr}(\rho_1(t),\rho_2(t))=\frac{1}{2}\Tr|\rho_1(t)-\rho_2(t)|,
\end{equation}
and the integral is evaluated over all time intervals where the trace distance increases $\partial_{t'}d_{Tr}(\rho_1(t'),\rho_2(t'))>0$. This measure has been used previously for a qubit in an evironment modeled by random matrices \cite{Pineda2019}.

For our system density matrix from Eq.~(\ref{evolr}), we prove in Appendix~\ref{app:XX_1q} that the measure $\mathcal{N}$ is nonzero only when $\kappa^2\leq 64 \alpha^2$. In this case, the measure is
\begin{equation}\label{Nonmarknew}
\mathcal{N}=\frac{1}{\exp\left(\frac{(\kappa+4\eta)\pi}{\sqrt{64\alpha^2-\kappa^2}}\right)-1}.
\end{equation}
Following the conclusion of Pang~\cite{Pang2017}, for no cooling term and no correction ($\kappa=\eta=0$), the measure $\mathcal{N}$ goes to infinity, indicating maximum non-Markovianity, as there is periodic information backflow from the bath to the system with no decay in amplitude. When $\kappa$ becomes nonzero, $\mathcal{N}$ becomes finite, indicating a weaker non-Markovianity due to the decay in the amplitude of the information backflow. Additionally, even with no cooling but with error correction, $\mathcal{N}$ can be positive.

If $\kappa^2 \geq 64 \alpha^2$, there is no increase in the trace distance between any two quantum states at any time, resulting in $\mathcal{N}=0$, indicating Markovian dynamics for this case.

In this section, we derived a Non-Markovian process via an interacting $X$-$X$ coupling between the system and a cooling bath, measured by a cooling rate $\kappa$. However, if the system's interaction with its environment is unknown or the source of noise is unspecified, a phenomenological approach is usually required.

\subsection{Post-Markovian master equation (PMME)}

We discussed how the dynamics of superconducting qubit systems can include non-Markovian effects. One could try to construct an explicit model for this non-Markovian environment that matches the observed behavior. However, it is often more practical to propose a memory kernel for the bath and benchmark its parameters to fit the data from the device \cite{Zhang2021,Agarwal2024}. In this subsection, we compare the Markovian case to the post-Markovian master equation (PMME) proposed in \cite{Shabani2005}, which includes the effect of past information via memory kernels. The PMME shown here can be solved analytically by Laplace transformation:
\begin{equation}\label{PMME}
	\frac{d}{dt}\rho(t)=\mathcal{L}_0\rho(t)+\mathcal{L}_1\int_{0}^{t}dt'k(t')\exp[(\mathcal{L}_0+\mathcal{L}_1)t']\rho(t-t')
\end{equation}
where $\rho(t)$ is the reduced density system state, $\mathcal{L}_0$ and $\mathcal{L}_1$ are generators of dissipative Markovian dynamics in Lindbald form, and $k(t)$ is a phenomenological memory kernel, which assigns weights to the history of previous system states. Note that an impulse function for kernel ($k_0(t)=\delta(t)$), which means no memory until time $t$, recovers a Markovian model of the form in Eq.~ (\ref{simplest}):
\[
\frac{d}{dt} \rho(t) = \mathcal{L}_0\rho(t) + \mathcal{L}_1\rho(t) = \eta \Gamma(\rho^S) + \gamma \mathcal{D}(\sigma_x) \rho^S .
\]

For our model, we use $\mathcal{L}_0=\eta \Gamma(\rho^S)$ as our error correcting generator and $\mathcal{L}_1=\gamma \mathcal{D}(\sigma_x)\rho^S$ as our error dissipator. Since we assume that the bath correlation gets weaker as the time runs, we consider two commonly proposed models for the memory kernel: an exponentially decaying kernel and an underdamped (or overdamped) exponential kernel. The parameters of the kernels depend on the bath correlation and could be solved for in an explicit bath model (like the spin-boson model \cite{Zihan2024}) or determined from explicit benchmarking data from the device \cite{Zhang2021}.

The first model---an exponentially decaying kernal with amplitude $a$ and decay rate $c$---is a simple case that allows closed analytical solutions. Its  Laplace transform is just a multiplicative inverse
\begin{equation}
	k_1(t)=ae^{-ct}\implies \tilde{k}_1(s)= \frac{a}{s+c}.
\end{equation}

For the second model we consider a kernel that represents an overdamped or underdamped situation:
\begin{equation}\label{kernel2}
	k_2(t) = \begin{cases}
		e^{-\frac{b}{2}t}\left(\frac{a-b/2}{\omega}\sinh(\omega t)+\cosh(\omega t)\right), & b^2\geq 4c, \\
		e^{-\frac{b}{2}t}\left(\frac{a-b/2}{\omega}\sin(\omega t)+\cos(\omega t)\right), & b^2<4c,
	\end{cases}
\end{equation}
where $a$ is the amplitude, $b/2$ is the decay rate, and the oscillating frequency is $\omega=\sqrt{|b^2-4c|}$. The kernel has the Laplace transform
\[
\tilde{k}_2(s)=\frac{s+a}{s^2+bs+c}.
\]

To analyze the one-qubit code in the PMME model, we start from Eq.~(\ref{PMME}) with a memory kernel $k(t)$, and the following dynamic generators: the bit flip Markovian generator $\mathcal{L}_1=\gamma \mathcal{D}(\sigma_x)$ from Eq.~(\ref{Xlindbaldian}) and the error correction generator $\mathcal{L}_0=\eta \Gamma^S$ from Eq.~(\ref{Gamma1qubit}). The derivation details can be found in Appendix~\ref{app:PMME_1q}.
Given the initial state in Eq.~(\ref{init}), the state at time $t$ is
\begin{eqnarray}\label{PMMEsOL}
	\rho(t) &=& \frac{1}{2}I + \frac{x_0}{2} e^{-\eta t} X + \frac{y_0}{2} \xi(t)Y \nonumber\\
    && + \left(\frac{z_0}{2}\xi(t)+\chi(t)\right)Z,
\end{eqnarray}
where the time-dependent amplitudes $\xi(t)$ and $\chi(t)$ are given by Eqs.~(\ref{xit}) and (\ref{chit}), respectively.

\subsubsection{Decaying exponential kernel}

At long times, for the exponential kernel $k(t)=ae^{-ct}$, the system behaves similarly to the previous cases in approaching a stationary state in which only diagonal terms of the density matrix (or ``populations'') survive:
\begin{equation}
\lim_{t\rightarrow \infty}\rho(t)=\frac{1}{2}I+\frac{1}{2}\frac{\eta(c^2-2a\gamma+c(2\gamma+\eta))}{c(2\gamma a+\eta(c+2\gamma+\eta))}Z.
\end{equation}
Again, for weak correction ($\eta\ll \gamma$) the state approaches the completely mixed state $\approx \frac{1}{2}I$, and for strong correction ($\eta \gg \gamma$) the state approaches $\frac{1}{2}I+\frac{1}{2}Z = \ket{0}\bra{0}.$

For the over- or underdamped kernels, the solution involves the roots of a cubic polynomial; while analytical solutions exists (in principle) they are complicated and not very illuminating, so we omit them here.

\section{Fidelity and the Zeno Effect}
\label{sec:Fidelity_ZE}

To evaluate the effectiveness of the error-correcting method, we use fidelity as a measure. Fidelity measures the closeness of two quantum states. Suppose the initial state of the coupled system and bath qubits is $\rho_{SB}(0)=\ket{0}\bra{0}_S\otimes \ket{0}\bra{0}_B$. The fidelity of the system’s density matrix with the initial state is
\begin{equation}\label{fidelity}
F(t) = \Tr\{(\ket{0}\bra{0}_S\otimes I_B)\rho_{SB}(t)\}.
\end{equation}
We will compute the system fidelity for our one-qubit density matrices for the three scenarios we have considered: Markovian bit flip decoherence, non-Markovian bit flip decoherence via a system-bath coupling with a cooling bath, and the PMME.

\subsection{One-qubit Markovian case with error correction}

\begin{figure}[h!]
	\centering
	\includegraphics[width=0.48\textwidth]{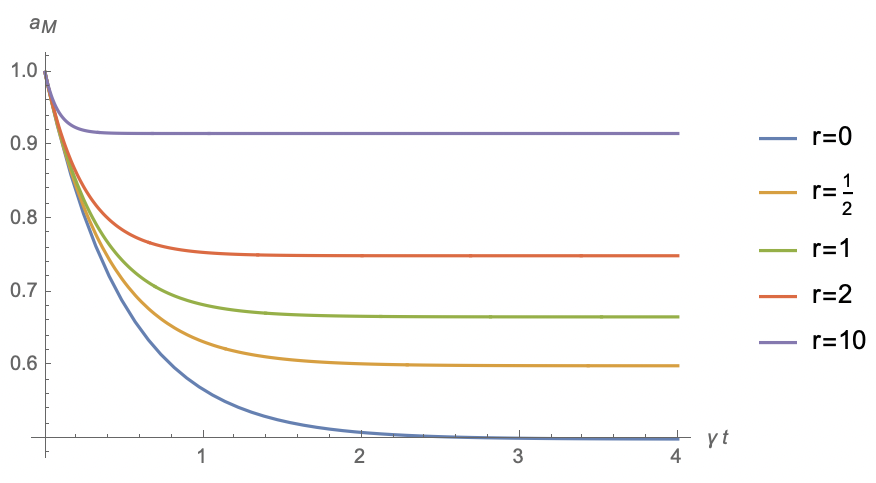}
	\caption{Single-qubit Markovian fidelity with continuous bit-flip errors and correction, as a function of time $\gamma t$ for different values of the ratio $r=\frac{\eta}{\gamma}$.}
 \label{fig:markovfidel2}
\end{figure}

We calculate the fidelity from Eq.~(\ref{fidelity}) by finding the solution of the Markovian master equation with error correction from Eq.~(\ref{ascend}) for the initial condition in Eq.~(\ref{init}). This gives us the following result:
\begin{equation}\label{aMarkov}
F_M(t) = \frac{\gamma+\eta}{2\gamma+\eta} + \frac{\gamma}{2\gamma+\eta} e^{-(2\gamma +\eta)t} .
\end{equation}
The plot of this solution is shown in Fig.~\ref{fig:markovfidel2}, where we note the monotonic decay to a constant asymptote in time. The stronger the error-correcting rate compared to the error rate, the closer the fidelity remains to 1, protecting the initial state.

The fidelity with bit-flip error rate $\gamma$ and correction rate $\eta$ decays initially at a rate $2\gamma+\eta$, and the limit at long times is
\begin{equation}
		F_{M\infty} = \lim_{t\rightarrow \infty} F_M(t) = \frac{\gamma+\eta}{2\gamma+\eta}.
\end{equation}
For small correction rate ($\eta\ll \gamma$) the fidelity approaches $\frac{1}{2}$, indicating a completely mixed final state, while for large correction rate ($\eta\gg \gamma$), the fidelity approaches $1$, indicating that the pure state $\ket{0}\bra{0}_S$ is preserved. The Markovian fidelity Eq.~(\ref{aMarkov}) can be written in terms of the stationary limit $F_{M\infty}$ as
\begin{equation}
		F_M(t) = F_{M\infty} + (1-F_{M\infty}) e^{-(2\gamma  +\eta)t}.
\end{equation}

\subsection{Non-Markovian system-bath $X$-$X$ coupling with a cooling bath}

We can obtain the fidelity in Eq.~(\ref{fidelity}) from the solution (\ref{evolr}) of the non-Markovian master equation with system-bath $X$-$X$ coupling with strength $\alpha$, a cooling bath with rate $\kappa$, and continuous error correction at rate $\eta$. Explicit expressions are given in Appendix~\ref{app:fidelity} Eq.~(\ref{nonmarkfidel}) for the non-Markovian case $\kappa^2<64\alpha^2$ and Eq.~(\ref{markfidel}) for the Markovian case $\kappa^2>64\alpha^2$. The fidelity decays at a rate $\eta+\kappa/4$, and the limit at long times is
\begin{equation}
		\lim_{t\rightarrow \infty} F(t)=\frac{\eta(\kappa +2\eta)+4\alpha^2}{\eta(\kappa +2\eta)+8\alpha^2}.
\end{equation}
 
\begin{figure}
		\centering
		\includegraphics[width=0.48\textwidth]{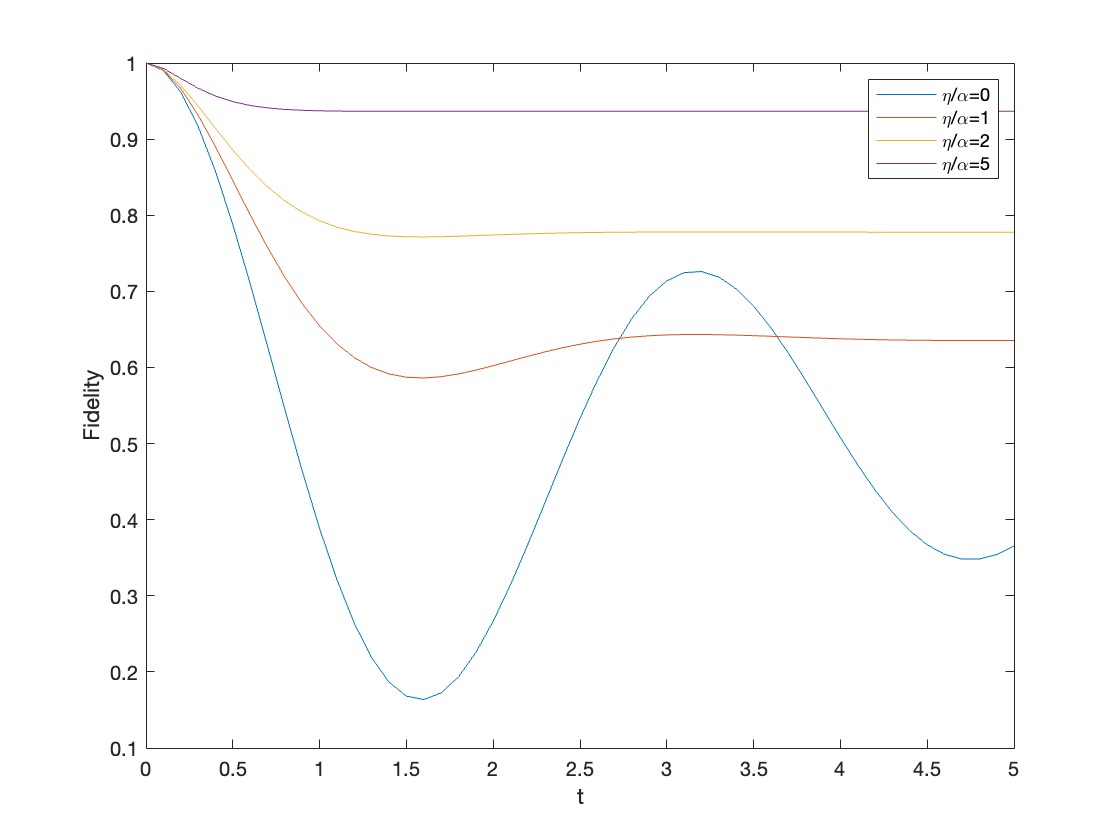}
	\caption{Single-qubit fidelity with $X$-$X$ system-bath coupling and error correction with fixed cooling rate  $\kappa = \alpha$, as a function of dimensionless time $\alpha t$, for different values of the error correcting rate $\eta$.}
 \label{fig:fidRkappa1NEW}
\end{figure}

In Fig.~\ref{fig:fidRkappa1NEW} the fidelity is shown with a non-zero cooling rate for several different correction rates. We observe that when there is weak error correction, oscillations occur, implying non-Markovian behavior. When the error-correction rate is increased, the oscillations become overdamped, and the system approaches its asympotic state (with fidelity closer to 1) faster. As shown in Eq.~(\ref{Nonmarknew}), higher error-correcting rates can induce Markovian behavior. We can compare this figure to the one shown in \cite{Oreshkov2007}, where there is no cooling procedure.

\begin{figure}[h!]
	\centering
	\includegraphics[width=0.45\textwidth]{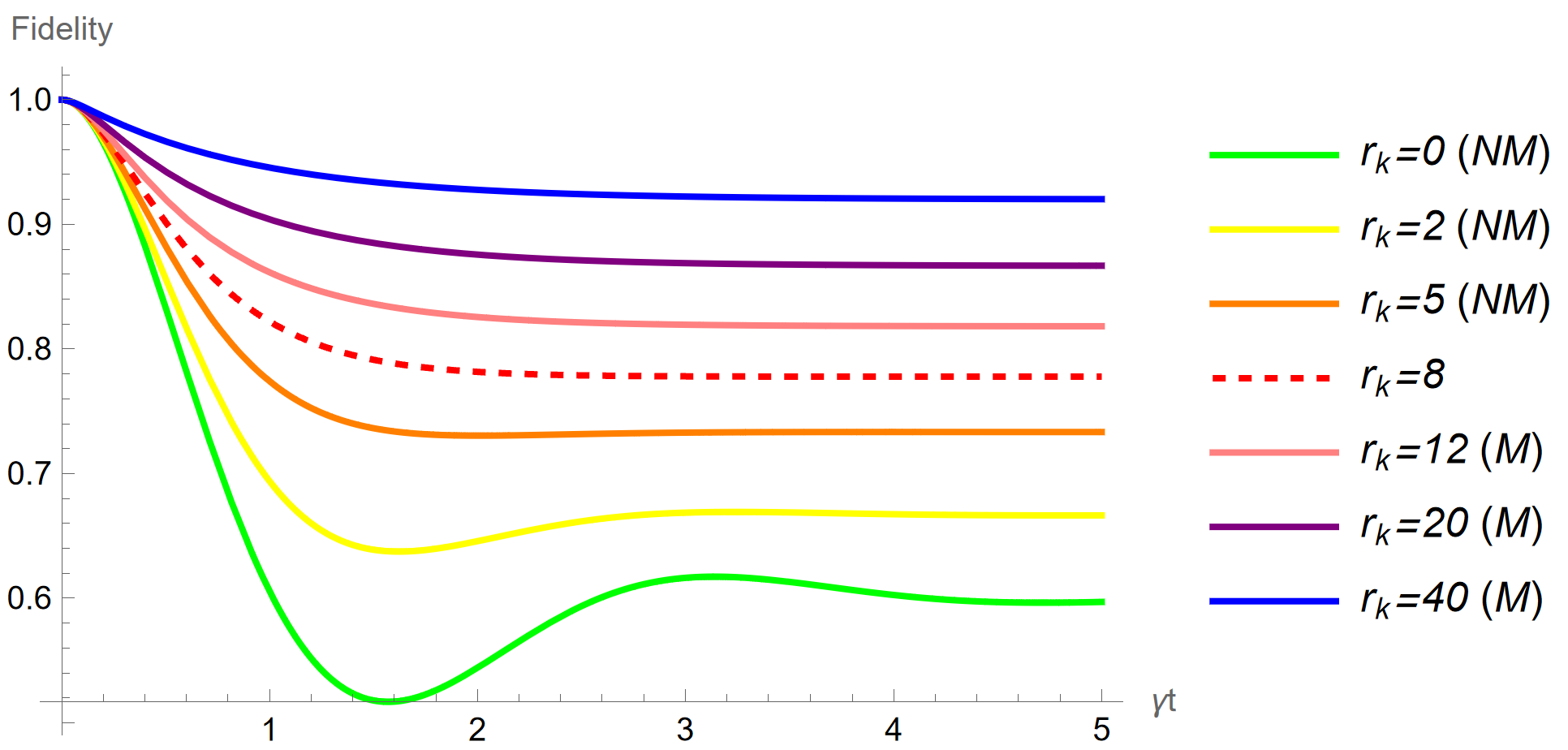}
	\caption{Single-qubit fidelity with $X$-$X$ system-bath coupling and fixed error-correcting rate $\eta=\alpha$, as a function of dimensionless time $\alpha t$ for different values of the cooling bath ratio $r_\kappa=\kappa/\alpha$. For $\kappa<8\alpha$ we have non-Markovian behavior, and for $\kappa\geq 8\alpha$ Markovian behavior.}
	\label{fig:fidkappafig}
\end{figure}

From Fig.~\ref{fig:fidkappafig}, representing the fidelity with fixed error-correcting rate and different cooling rates, we can observe two different behaviors. In the non-Markovian case ($\kappa^2<64\alpha^2$), the fidelity oscillates while decaying until reaching its asymptotic value. In the Markovian case ($\kappa^2\geq 64\alpha^2$), the fidelity monotonically decreases towards the asymptotic value. In addition, we notice that the higher the cooling rate, the higher the fidelity, so the cooling process also acts as a kind of ``correction'' \cite{Sarovarcool}.

\subsection{Post-Markovian Master Equation (PMME)}

The fidelity of the PMME for the one-qubit model, whose solution is given in Eq.~(\ref{PMMEsOL}), is
\begin{equation}
F(t) = \frac{1}{2}(1+\xi(t))+\chi(t)) ,
\end{equation}
where $\xi(t))$ and $\chi$ are given in Eqs.~(\ref{xit}) and (\ref{chit}), respectively. A simple analytical result for the exponential kernel is available, while the overdamped solution requires solving a cubic equation to obtain the roots in the Laplace transform.

\subsubsection{Decaying exponential kernel}

\begin{figure}[h!]
	\centering
	\includegraphics[width=0.45\textwidth]{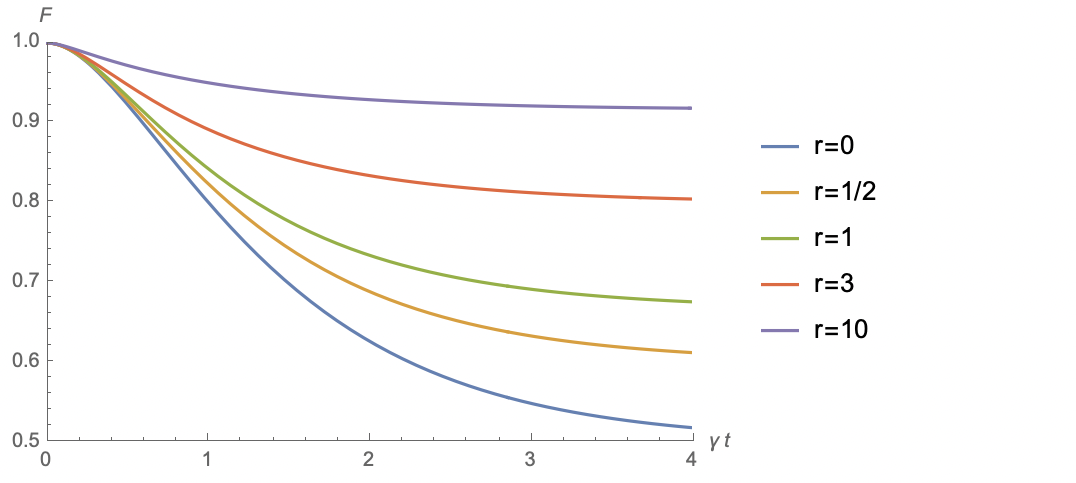}
	\caption{Single-qubit PMME solution fidelity with exponential kernel $k_1(t)=ae^{-ct}$, where $a=1$ and $c=1$, as a function of dimensionless time $\gamma t$ for different values of the error-correcting ratio $r=\eta/\gamma$.}
 \label{fig:correcta}
\end{figure}

In Fig.~\ref{fig:correcta} we show the fidelity of the one-qubit PMME for an exponential kernel with fixed parameters of amplitude and decay and different correction rates. Note that stronger correction results in fidelities closer to 1. At short times, we note the convex parabolic decay due to the Zeno effect, followed by a monotonic decay at long times.

\subsubsection{Decaying oscillating damped exponential kernel}

\begin{figure}[h!]
		\centering
	\includegraphics[width=0.45\textwidth]{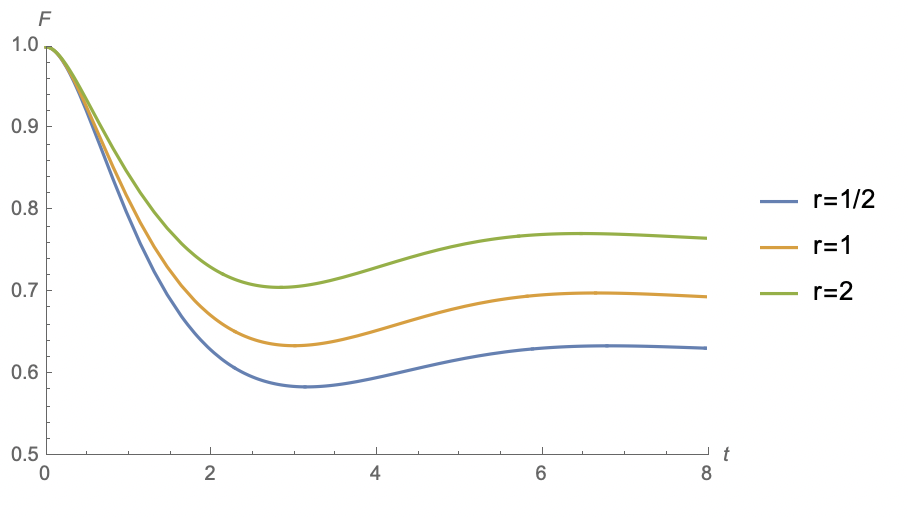}
		\caption{Single-qubit PMME solution fidelity with oscillating exponential kernel $k_2(t)=e^{-\frac{b}{2}t}\left(\frac{a-b/2}{\omega}\sin(\omega t)+\cos(\omega t)\right)$, as a function of dimensionless time $\gamma t$, for different values of the ratio $r=\eta/\gamma$. Here, $a=b=c=1$.}
  \label{plotk2fig}
	\end{figure}

In Fig.~\ref{plotk2fig} we show the fidelity as a function of time for the model in Eq.~(\ref{kernel2}), where the kernel simulates an underdamped case. The Laplace transform of the kernel $\tilde{k}_2=(s+a)/(s^2+bs+c)$ depends on three parameters, which we fix to be $a=b=c=1$ in arbitrary units defined by the decay rate $\gamma$.

In contrast to the PMME with exponentially decaying kernel, the damped kernel exhibits characteristic parabolic decay at short times due to the Quantum Zeno Effect. At longer times, oscillations occur that are not observed in the exponential kernel.

\subsection{Quantum Zeno Effect}

The Quantum Zeno Effect is a phenomenon where the time evolution of a quantum system's state is ``frozen'' or slowed down by applying frequent measurements, nearly continuously. While actually freezing the evolution would require the measurement rate to diverge, one can see slowing even at finite measurement rates.

\subsubsection{Short-time behavior}

In our case, the Zeno effect appears in an open system when initially the system and the environment are uncorrelated, but they interact over time via a Hamiltonian \cite{BrunQEC}. This interaction results in the system state having a quadratic dependence on time during its initial moments.

For instance, the one-qubit Markovian fidelity in Eq.~(\ref{aMarkov}), with bit flip errors and correction, can be expanded at short times $t$ as
\begin{equation}
F_M(t)\approx 1-\gamma t+\frac{1}{2}(2\gamma+\eta)\gamma t^2+O(t^3),
\end{equation}
showing a non-zero linear term in time. This is exponential decay, indicating no Zeno effect.

On the other hand, the fidelity of the $X$-$X$ coupled model in Eq.~(\ref{nonmarkfidel}) exhibits a quadratic dependence on time, characteristic of the Zeno effect:
\begin{equation}\label{eq:fidelityshortXX1q}
	F_{XX}(t)\approx 1-\alpha^2 t^2+\frac{1}{6}(\kappa+4\eta)\alpha^2 t^3+O(t^4) .
\end{equation}

The fidelity for the one-qubit PMME with a decaying exponential kernel for short times is
\begin{equation}
F_{PMME}(t)\approx 1 - \frac {1}{2} a \gamma t^2 + O(t^3) ,
\end{equation}
which shows quadratic decay at short times, indicative of the Zeno effect. Note that the coefficient of the $t^2$ term depends on the amplitude of the kernel $a$ and the error rate $\gamma$, but not on the decay rate $c$ of the kernel.

\subsubsection{Long-time behavior}

We now analyze the asymptotic one-qubit infidelity, the weight outside the code space:
\[
1-F_{\infty}=\lim_{t\rightarrow\infty}(1-F(t)) ,
\]
for each of the three models. In the Markovian case, Eq.~(\ref{aMarkov}), with bit-flip errors and correction, we can expand in the inverse of error correction factor $1/\eta$ to obtain
\begin{equation}
1-F_{M\infty}=\frac{\gamma}{2\gamma+\eta}\approx \frac{\gamma}{\eta}-2\left( \frac{\gamma}{\eta}\right)^2+O\left( \frac{1}{\eta^3}\right).
\end{equation}

The asymptotic infidelity of the $X$-$X$ coupled system-bath with cooling bath and correction, Eq.~(\ref{nonmarkfidel}), is quadratic to leading order in $1/\eta$:
\begin{equation}
	1-F_{XX\infty}=\frac{4\alpha^2}{\eta(\kappa +2\eta)+8\alpha^2}\approx4\left(\frac{\alpha}{\eta}\right)^2-8\frac{\alpha^2\kappa}{\eta^3}+O\left( \frac{1}{\eta^4}\right).
\end{equation}

For the PMME model with exponential memory kernel $ae^{-ct}$, the infidelity at long times is
\begin{equation}
	1-F_{PMME\infty}=\frac{a\gamma (c+\eta)}{c(2a\gamma+\eta(2\gamma+c+\eta))}\\
 \approx\frac{a}{c}\frac{\gamma}{\eta}+O\left(\frac{1}{\eta^2}\right).
\end{equation}
This shows that the infidelity decays to first order as the inverse of the correction rate $1/\eta$, and the coefficient depends on both parameters of the memory kernel (amplitude $a$ and decay rate $c$).

\section{More general  $[[n,k]]$ codes }
\label{sec:More_general_codes}

For an $[[n,k]]$ code that protects $k$ logical qubits by encoding them into $n$ physical qubits, we can define a decoding map $\mathfrak{D}(\rho)$ that goes from the encoded basis with $n$ qubits to the decoded basis with $k$ qubits. This map can take the form
\begin{equation}
	\mathfrak{D}(\rho)=\Tr_{anc}\left\{U_{\mathcal{E}}^{\dagger}\left(\sum_s C_sP_s\rho P_sC_s^{\dagger}\right) U_{\mathcal{E}}\right\} ,
\end{equation}
where the partial trace is taken over the $n-k$ ancilla qubits, the sum is done over the error syndromes $s$, the syndrome projectors $P_s$ project onto the error space corresponding to syndrome $s$ while the operators $C_s$ are the unitary corrections for each syndrome. The operator $U_{\mathcal{E}}$ is the unitary encoding operator that maps the $k$ qubits (plus $n-k$ ancillas in the state $\ket0$) to the codeword on $n$ physical qubits. Therefore, the error-correcting generator is
\[
\Gamma(\rho)=\Phi(\rho)-\rho ,
\]
where the associated CPTP (completely positive trace-preserving) map $\Phi(\rho)$ is
\begin{equation}\label{eq:CPTP_error}
	 \Phi(\rho) = \sum_s C_sP_s\rho P_sC_s^{\dagger}.
\end{equation} 
Hence, the error-correcting method can be viewed both as a quantum-jump process, as proposed by Zurek \cite{Zurek1998}, with a full error-correcting operation applied at random times with rate $\eta$, and as a continuous sequence of infinitesimal CPTP maps \cite{Oreshkov2007}
\[
\rho \rightarrow (1-\eta dt)\rho + \eta dt \Phi(\rho) .
\]
This form can arise as the continuous limit of a sequence of weak measurements followed by weak unitary corrections \cite{Hsu2016}.

\subsection{The three-qubit code}

The simplest $[[n,k]]$ stabilizer code that has a nontrivial code space is the three-qubit code with $n = 3$ physical qubits protecting $k = 1$ qubit of information against bit-flip errors affecting any one of the three qubits. The quantum three-qubit code is the quantum version of the classical repetition code for $n=3$, and inherits the majority vote decoding algorithm from its classical version. However, it neither detects phase errors nor corrects bit-flip errors occurring on multiple qubits.

In Appendix~\ref{app:3qcode}, we describe in detail the encoding unitary, the four projection operators, and their corresponding unitary corrections for each error syndrome. The error-correcting generator is 
\begin{equation}\label{eq:3bit_error_gen}
	\begin{split}
		\Gamma(\rho)=&(\ket{000}\!\bra{000}+\ket{111}\!\bra{111})(\rho+X_1\rho X_1+X_2\rho X_2\\
  &+X_3\rho X_3)(\ket{000}\!\bra{000}+\ket{111}\!\bra{111})-\rho,
	\end{split}
\end{equation} 
where $X_i=(\sigma_x)_i=I\otimes...\otimes X\otimes...\otimes I$ represents the Pauli matrix $X$ acting on qubit $i$. To avoid writing many tensor products we will, e.g., denote $X\otimes I \otimes X$ as $X_1 X_3$.

\subsubsection{Markovian master equation for the three-qubit code}

\begin{figure}[h!]
\centering
\includegraphics[width=0.3\textwidth]{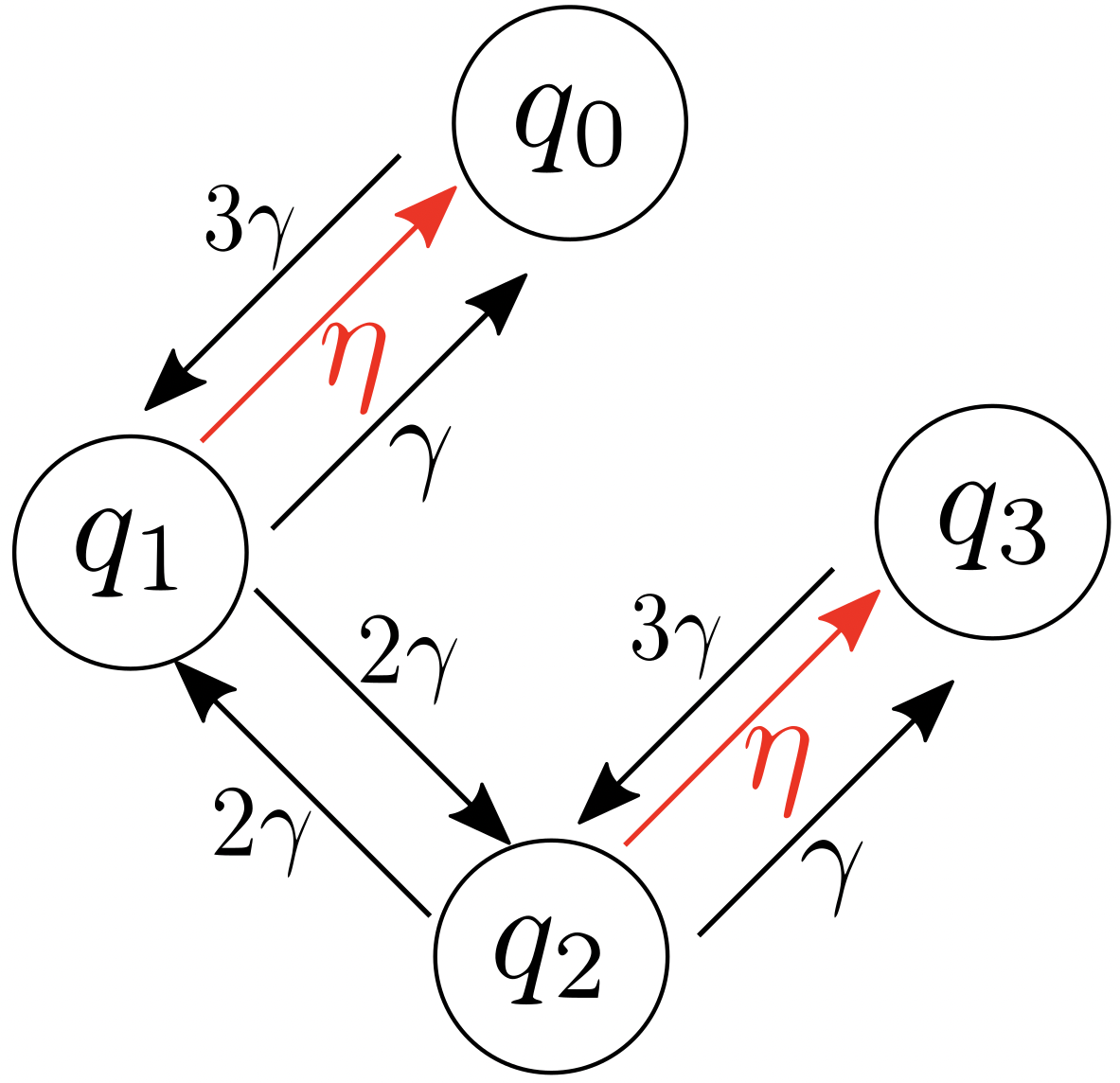}
\caption{Chain diagram for the three qubit-code with Markovian bit-flip noise and continuous correction, which shows the allowed transitions among the coefficients with rates $\gamma$ for bit-flip errors (in black) and error corrections at rate $\eta$ (in red).}
\end{figure}

The Markovian master equation for the three-qubit code with bit-flip noise at rate $\gamma$ and continuous correction at rate $\eta$ is
\begin{equation}
\partial_t \rho^S = \gamma \mathcal{D}(\rho)
+ \eta \Gamma(\rho), 
\end{equation}
where we assume that a bit-flip error can occur in any of the physical qubits with equal probability. The bit-flip Lindbladian is
\begin{eqnarray}
\mathcal{D}(\rho) &=& \sum_{i=1}^{3}\left(X_i\rho X_i-\rho\right)\hspace{-0.1cm} \nonumber\\
&=& X_1 \rho X_1+X_2\rho X_2+X_3 \rho X_3-3\rho
\end{eqnarray} 
The solution for the master equation can be expressed as a linear combination
\begin{equation}\label{rho3qubit}
\rho(t) = q_0(t)\rho(0) + q_1(t)\rho_1 + q_2(t)\rho_2 + q_3(t)\rho_3
\end{equation} 
where $\rho_i$ denotes the component of the state with Hamming error weight $i$:
\begin{equation}
\begin{split}
\rho_1 &= \frac{1}{3}(X_1\rho(0)X_1 + X_2\rho(0)X_2 + X_3\rho(0)X_3),\\
\rho_2 &= \frac{1}{3}(X_1X_2\rho(0)X_1X_2 + X_2X_3\rho(0)X_2X_3 \\
  &\qquad+X_1X_3\rho(0)X_1X_3),\\
\rho_3 &= X_1X_2X_3\rho(0)X_1X_2X_3 .
	\end{split}
\end{equation} 
Each is an equally-weighted mixture of single-qubit, two-qubit and three-qubit errors on the initial state. The explicit time-dependent coefficients are given in Eq.~(\ref{3markov}).

\begin{figure}[h!]
\centering
\includegraphics[width=0.45\textwidth]{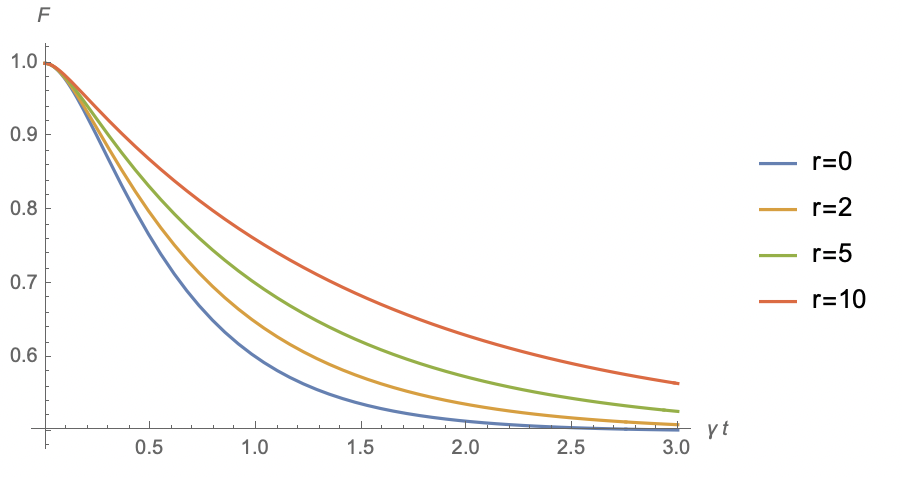}
\caption{Overlap fidelity $q_0(t)+q_1(t)$ for the three-qubit code with Markovian noise as a function of dimensionless time $\gamma t$ for different values of the ratio $r=\eta/\gamma$.}
\label{3qmarkovnaruto}
\end{figure}

\begin{figure}
\centering
\includegraphics[width=0.45\textwidth]{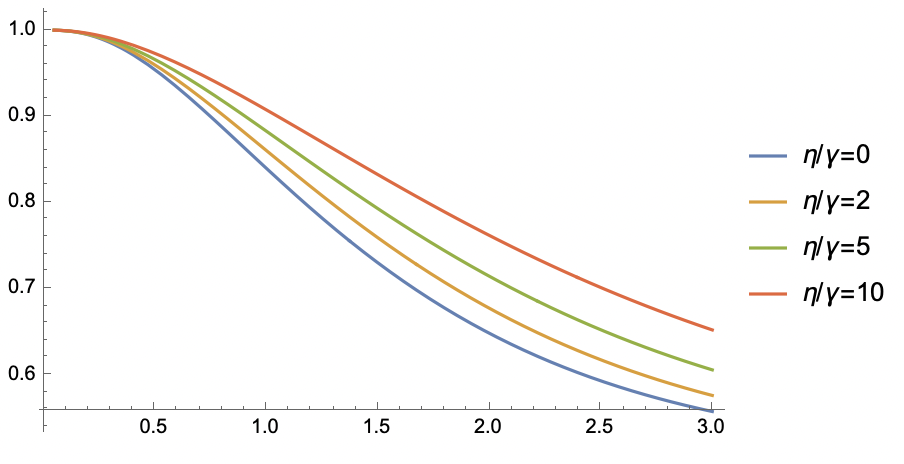}
\caption{Overlap fidelity $q_0(t)+q_1(t)$ for the three-qubit code with non-Markovian noise in the PMME model as a function of the dimensionless time $\gamma t$. Here we used the exponential kernel $k(t)=e^{-t}$ for different values of the ratio $\eta/\gamma$.}
\label{fig:pmme3qerror}
\end{figure}

The master equation
\[
\partial_t \rho = -i[H, \rho] + \gamma\mathcal{D}(\rho) + \eta \Gamma^S(\rho)
\]
is solved by constructing the superoperator matrix and taking its exponential, as shown in Appendix~\ref{app:3qcode}.

In the limit where the error correction rate is much larger than the dissipation rate, $\eta \gg \gamma$, we recover 
\begin{equation}
\lim_{t \rightarrow \infty} \rho^S(t) = \frac{1}{2} \left(\rho(0)+X_1X_2X_3\rho(0)X_1X_2X_3\right).
\end{equation}
The difference from the single-qubit model is that in this code there are \textit{logical} errors that allow the decay of the state’s fidelity inside the code space. The overlap fidelity $F(t) = q_0(t)+q_1(t)$ shown as a function of time in Fig.~\ref{3qmarkovnaruto}, at short times decays as 
\begin{equation}\label{markov3t2}
F_M(t) \approx 1 - 3(\gamma t)^2 + \gamma^2(8\gamma+\eta)t^3 + O(t^4).
\end{equation}

\subsubsection{PMME for the three qubit code}

\begin{figure}
	\centering
	\includegraphics[width=0.45\textwidth]{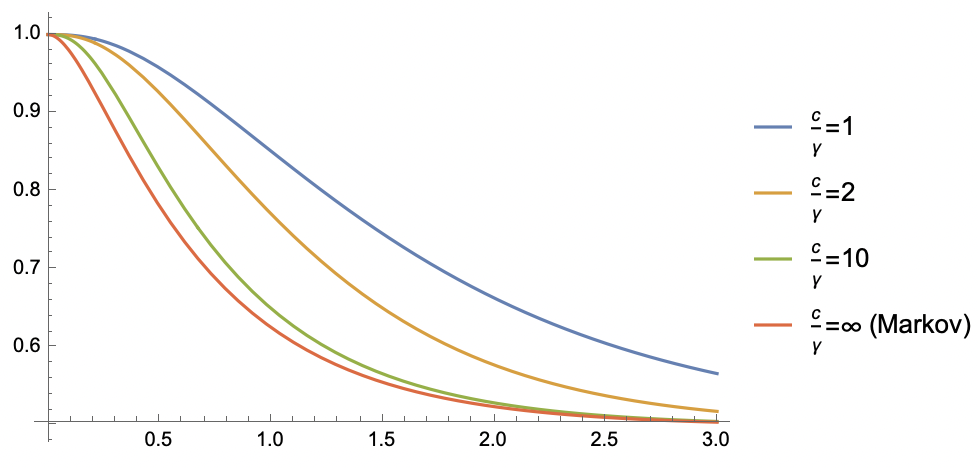}
\caption{Three-qubit overlap fidelity for PMME with an exponential kernel $ce^{-ct}$ for different values of $c$ as a function of dimensionless time $\gamma t$ with error-correction rate $\eta=\gamma$.}
\label{figkerc}
\end{figure}

We now generalize the previous Markovian case to the post-Markovian master equation (PMME) as described in Eq.~(\ref{PMME}). It is easiest to find an analytical solution with an exponential kernel where the amplitude and decay rate are equal, $k(t)=ce^{-ct}$. In the limit $c\rightarrow\infty$ the kernel approaches the Dirac delta distribution, which is associated with the Markovian limit. In Fig.~\ref{figkerc}, we show the overlap fidelity for PMME with the exponential kernel $ce^{-ct}$ with constant error correction at same rate as the error rate, $\eta=\gamma$, for different values of $c$. We observe that smaller $c$ results in a stronger Zeno effect, while in the limit of large $c$ we recover the Markov case shown in Fig.~\ref{3qmarkovnaruto}.
 
The corresponding expression for the overlap fidelity is given in Eq.~(\ref{fidpmme3qexp}), and it is plotted as a function in time for different error-correcting rates $\eta$ in Fig.~\ref{fig:pmme3qerror}. At short times, the fidelity can be expanded as
\begin{equation}\label{fidpmme3q}
F_{PMME}(t)= 1-c \gamma^2 t^3 +\frac{1}{4}c\gamma^2 (c+8\gamma+\eta)t^4+O(t^5),
\end{equation}
and observe that the first significant time-dependent term is cubic rather than quadratic (as in the Markovian case). Hence, the non-Markovian fidelity decays more slowly in time compared to its Markovian equivalent.

\subsubsection{Three system and three bath qubits with $X$-$X$ coupling}

\begin{figure}
\begin{subfigure}{.45\textwidth}
  \centering \hspace{-1cm}
  \includegraphics[width=1\linewidth]{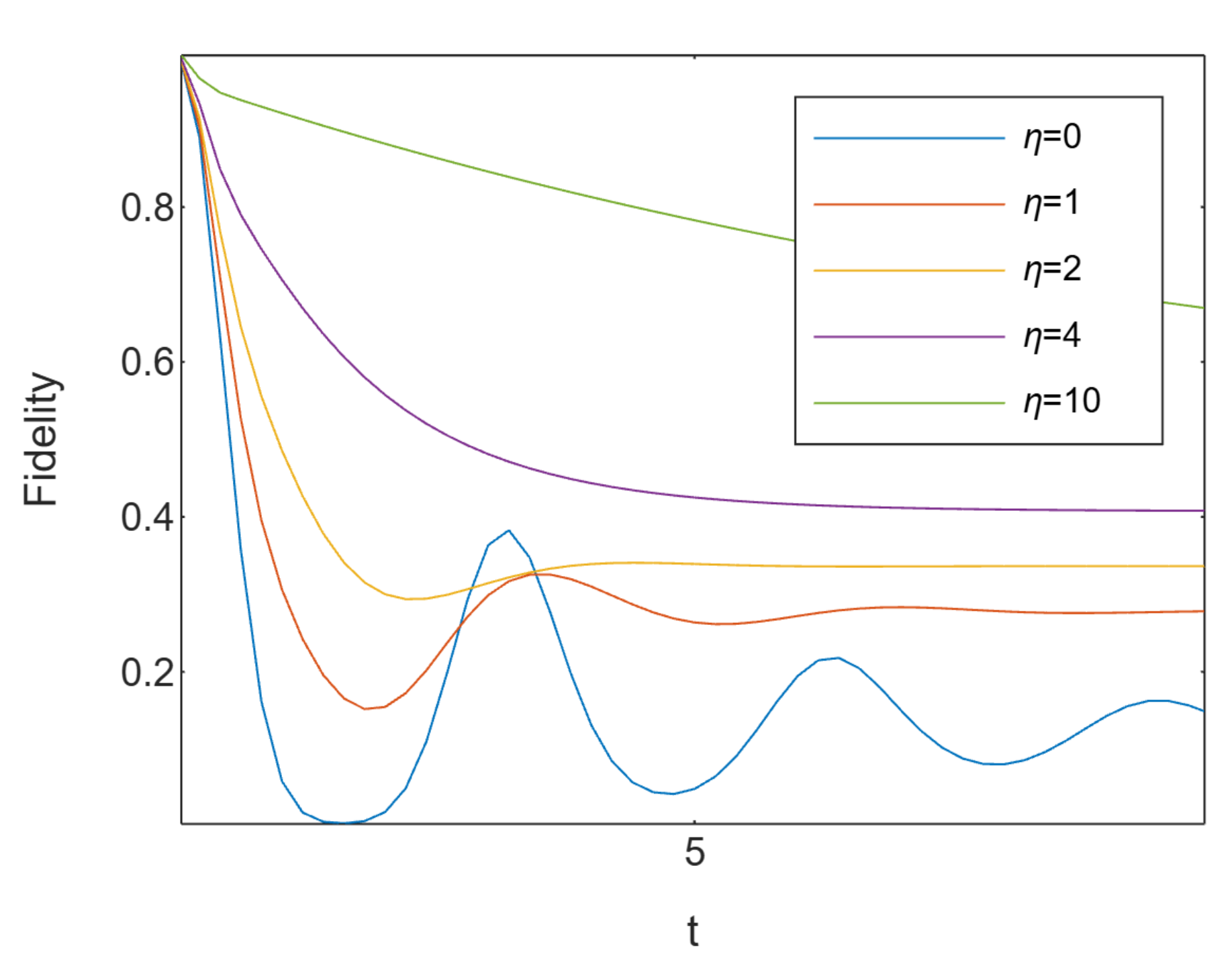}
  \caption{$\kappa = \alpha$.}
  \label{fig:traced_J3_a05}
\end{subfigure}%
   \vskip\baselineskip
\begin{subfigure}{.45\textwidth}
\vspace{-0.3cm}
  \centering 
\includegraphics[width=1\linewidth]{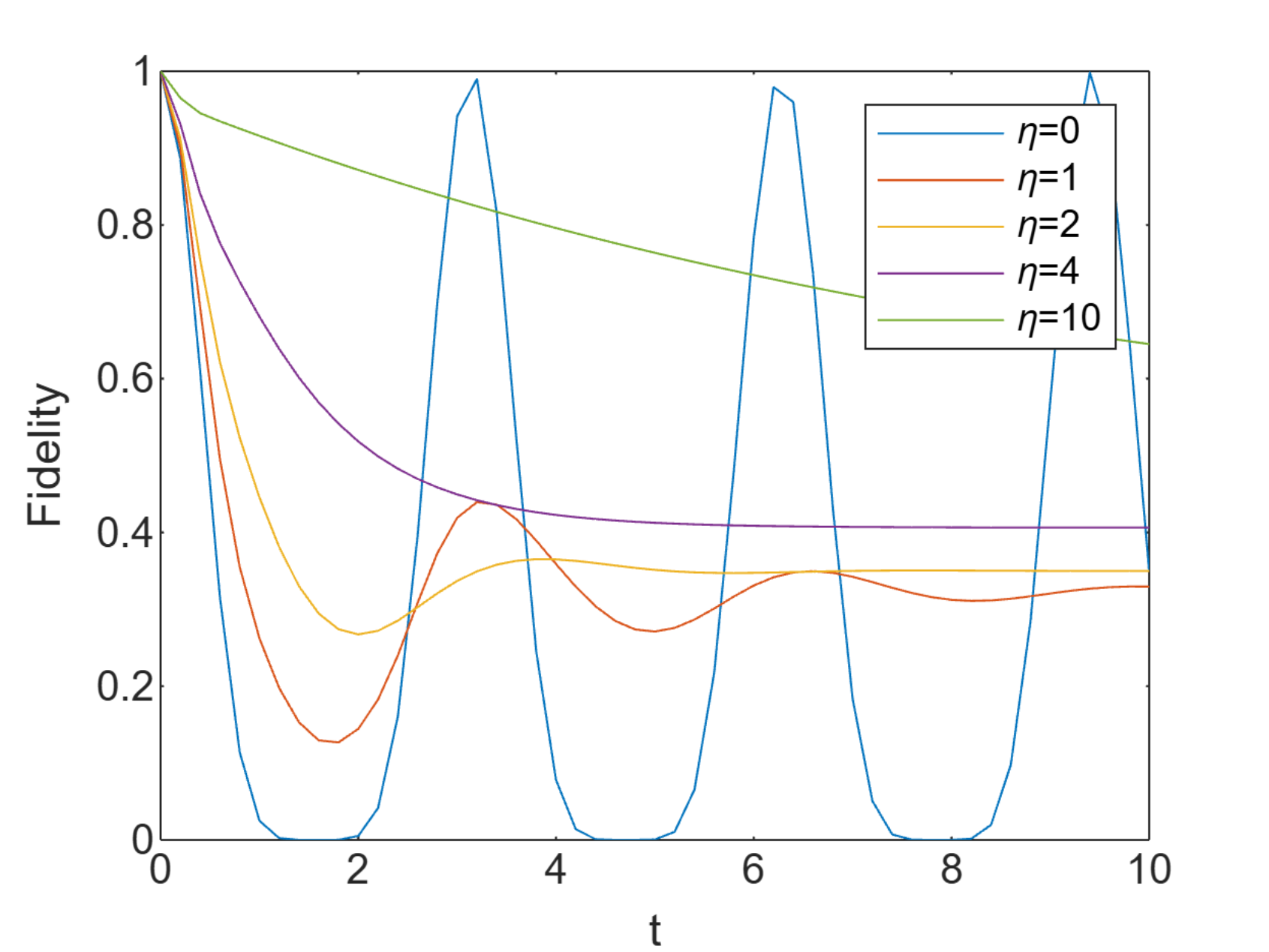}
  \caption{$\kappa = 0$.}
\end{subfigure}%
   \caption{Three-qubit fidelity with bit flip system-bath coupling and error correction (a) with cooling bath or (b) without cooling, as a function of dimensionless time $\alpha t$ for different values of the error correcting rate $\eta$.}
   \label{fig:3qubit-XXcoup-constk}
\end{figure}

\begin{figure}
	\centering
	\includegraphics[width=0.45\textwidth]{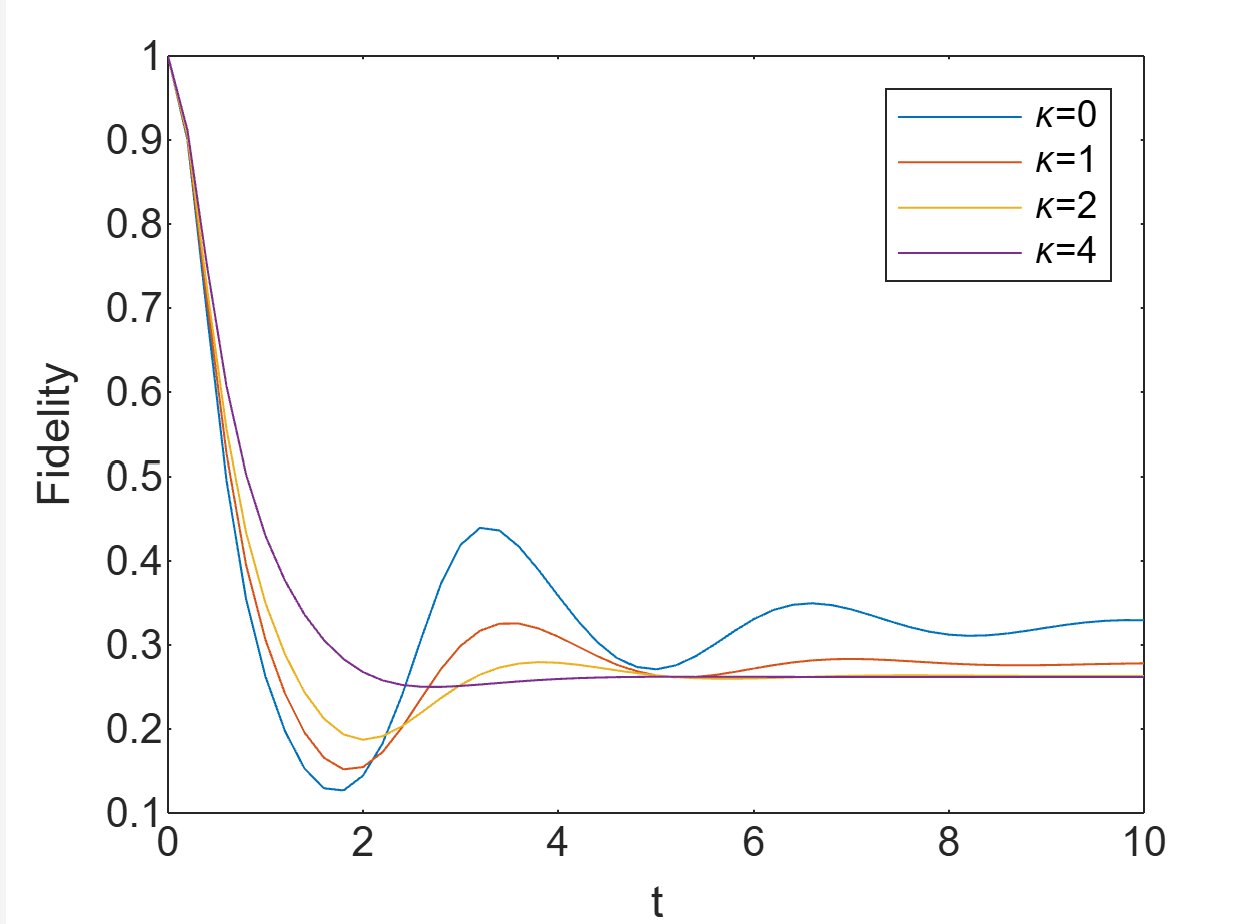}
\caption{Three-qubit fidelity with bit-flip system-bath coupling and error correction at rate $\eta = \alpha$ for different cooling rates $\kappa$, as a function of dimensionless time $\alpha t$.}
\label{fig:3qubit-diffk}
\end{figure}

To generalize the $X$-$X$ coupling of the single-qubit code to the three-qubit repetition code, we will need a model with six qubits: three for the system (used for the codeword) and three for the environment. We assume that each system qubit is coupled to its respective bath qubit with the same coupling strength,
\begin{equation}\label{eq:XX_3qubit}
    H=\alpha \sum_{i=1}^3 X_i^S \otimes X_i^B ,
\end{equation}
where the superscript $S$ indicates that it is acting on one of the system qubits and $B$ indicates acting on the respective bath qubit. While this Hamiltonian acts on both system and bath, the cooling process acts only on the bath qubits, while the error-correcting procedure acts only on the system qubits.

Assuming that the cooling rate $\kappa$ is the same for all the bath qubits in the master equation, then
\begin{equation}\label{eq:dissip3qubit}
    \kappa \mathcal{D}(\rho)= \kappa \sum_{i=1}^3 \mathcal{D}[\sigma_{i-}^B](\rho) ,
\end{equation}
where each $\mathcal{D}[\sigma_{i-}^B]$ acts as in Eq.~(\ref{eq:cool_disip}) only on its respective bath qubit and as the identity on the other qubits.

In Fig.~\ref{fig:3qubit-XXcoup-constk} we show the numerical dynamics of the fidelity for the three-qubit code with $X$-$X$ system-bath coupling and error correction, for both the case with cooling ($\kappa = \alpha$) and without cooling ($\kappa = 0$), for different correction rates $\eta$. When there is no cooling and no correction, the fidelity oscillates with period $T=\pi/\alpha$; however, where cooling and correction are present we recover asymptotic decay in the long run. In both scenarios, stronger correction rates yield higher fidelity. At very short times we note the concavity of the curve, indicating the Zeno effect. 

In Fig.~\ref{fig:3qubit-diffk} we show the dynamics of the fidelity for the three-qubit code with $X$-$X$ system-bath coupling and error correction for different cooling rates $\kappa$, similar to its one-qubit counterpart in Fig.~\ref{fig:fidkappafig}. Note that for small cooling rate the fidelity has non-Markovian oscillations, while for higher rates, the fidelity experiences a monotonic (almost exponential) decay.

The short-time fidelity, using the Taylor expansion, results in
\begin{equation}\label{fidxx3q}
F_{XX}(t)=1-3\alpha^2 t^2+\frac{1}{2}\alpha^2(\kappa+2\eta)t^3+O(t^4) .
\end{equation}
As one can observe, the first significant time-dependent term is quadratic, just as in the Markovian case in Eq.~(\ref{markov3t2}), indicating that the Zeno effect does not suppress the leading order in the Fidelity decay for the $X$-$X$ coupling model. In fact, it has the same behavior as the short-time fidelity of the one-qubit case in Eq.~(\ref{eq:fidelityshortXX1q}), except with the coefficient multiplied by three.

\subsection{The five-qubit $[[5,1,3]]$ code}

As previously mentioned, the three-qubit code only corrects weight-one bit-flip errors (or, using the code space in its Hadamard basis, only weight-one phase-flip errors). However, both $X$ and $Z$ errors aren't corrected simultaneously. In fact, no quantum code that can protect against an arbitrary single-qubit error can be shorter than 5 qubits \cite{Brun2019,Knill2001}. The $[[5,1,3]]$ code, discovered independently by Bennett et al \cite{Bennett1996} and Laflamme et al. \cite{Laflamme1996}, is a non-CSS code: a stabilizer code that cannot be written as the intersection of two classical codes for correcting $Z$ and $X$ Pauli errors, respectively. Since the code uses $n=5$ physical qubits to encode $k=1$ logical qubit, we need $n-k=4$ independent stabilizer generators: 
\begin{equation}\label{eq:stabs_five}
    \begin{split}
&S_1=XZZXI,\quad S_2=IXZZX,\\
&S_3=XIXZZ,\quad S_4=ZXIXZ.
    \end{split}
\end{equation}
The logical operators can be written
\begin{equation}\label{eq:logicals_five}
\bar{X}=XXXXX,\bar{Z}=ZZZZZ.
\end{equation}
The logical operators commute with all the stabilizer elements, and take states in the code space to other states in the code space. They can be thought of as encoded Pauli gates, or alternatively as logical Pauli errors (since the stabilizers cannot even detect these errors).

The product of a stabilizer and a logical operator will still be a logical operator, so logical operators can be written in different forms. For example, the logical operator $\bar{X}$ could be also written as $S_1\bar{X} = IYYIX$. Hence, this logical operator can be written as a weight-three operator, indicating that the code distance is $d=3$.

The error-correction procedure consists in applying a Pauli correction corresponding to the measured syndrome (the binary pattern of commutation and anticommutation with the stabilizers), as shown in Table~\ref{tab:5qubitsyndromes}. Here. $0$ indicates that the error commutes with the corresponding stabilizer generator (which will be measured to be $+1$), and $1$ that it anticommutes with that stabilizer generator (which will be measured to be $-1$). This table allows one to construct the explicit projectors $P_s$ and correcting operators $C_s$ for a given error syndrome $s$, as in Eq.~(\ref{eq:CPTP_error}). We use this to construct the error-correcting generator $\Gamma(\rho)$ in matrix form in Appendix~\ref{app:five_qubit_code}.

\begin{table}[h!]
\begin{center}
\begin{tabular}{
|c|c c c c|c  |c|c c c c| }
 \hline
 Error & $S_1$ & $S_2$ & $S_3$ & $S_4$ &\quad \quad &Error & $S_1$& $S_2$ & $S_3$ & $S_4$   \\
 \hline
$I$ & 0 & 0 & 0 & 0&\quad\quad &$Z_3$ & 0 & 0 & 1 & 0\\
$X_1$ & 0 & 0 & 0 & 1 &\quad \quad & $Z_4$ & 1 & 0 & 0 & 1\\
$X_2$ & 1 & 0 & 0 & 0 &\quad \quad & $Z_5$ & 0 & 1 & 0 & 0 \\
$X_3$ & 1 & 1 & 0 & 0 &\quad \quad &$Y_1$ & 1 & 0 & 1 & 1\\
$X_4$ & 0 & 1 & 1 & 0 &\quad \quad &$Y_2$ & 1 & 1 & 0 & 1\\
$X_5$ & 0 & 0 & 1 & 1 &\quad \quad &$Y_3$ & 1 & 1 & 1 & 0\\
$Z_1$ & 1 & 0 & 1 & 0 &\quad \quad &$Y_4$ & 1 & 1 & 1 & 1 \\
$Z_2$ & 0 & 1 & 0 & 1 &\quad \quad& $Y_5$ & 0 & 1 & 1 & 1\\
\hline
\end{tabular}
\end{center}
\caption{Error syndromes for all the single-qubit Pauli errors for the $[[5,1,3]]$ quantum error-correcting code.}
\label{tab:5qubitsyndromes}
\end{table}

\subsubsection{Markovian master equation for the five-qubit code}

Again, we express the density matrix as a linear combination of density matrices with errors of different Hamming weight:
\begin{equation}
\begin{split}
\rho(t)&=q_0(t)\rho_0+q_1(t)\rho_1+\sum_{i=2}^{5}\sum_{j=1}^iq_{ij}(t)\rho_{ij}\\
&=q_0\rho_0+q_1\rho_1+q_{2A}\rho_{2A}+q_{2B}\rho_{2B}+q_{3A}\rho_{3A}\\
&+q_{3B}\rho_{3B}+q_{3C}\rho_{3C}+q_{4A}\rho_{4A}+q_{4B}\rho_{4B}+q_{4C}\rho_{4C}\\
&+q_{4D}\rho_{4D}+q_{5A}\rho_{5A}+q_{5B}\rho_{5B}+q_{5C}\rho_{5C}+q_{5D}\rho_{5D}\\
&+q_{4E}\rho_{5E} ,
\end{split}
\end{equation}
where each of the subdensity matrices are equally weighted mixtures of Pauli errors on the initial state:
\begin{equation}\label{eq:subdensity_5}
    \rho_{ij}=\frac{1}{n}\sum_{k=1}^{n}E_{k}\rho(0)E_{k} .
\end{equation}
The error operators are listed in Table~\ref{tab:5qubitgroups}. From the table we notice that while the stabilizer generators in Eq.~(\ref{eq:stabs_five}) correspond to $\rho_{4A}$, the logical errors correspond to $\rho_{3B}$, $\rho_{5C}$ and $\rho_{5E}$.

\begin{table}
\begin{center}
\begin{tabular}{|c|c|c|c|}
\hline
 & Type &Example & $n$ \\
 \hline
 \hline
$0$ & $I^{\otimes 5}$& $IIIII$ & 1 \\
 \hline
 $1$ & $\sigma^a_i$&  $X_1=XIIII,Y_3=IIYII$ & 15 \\ 
 \hline
 $2A$ & $\sigma^a_i\sigma^a_j$&  $XXIII,IIYIY,ZIIIZ$ & 30 \\ 
 $2B$ & $\sigma^a_i\sigma^b_j$&  $IIIXY,IYIIZ,XIIZI$ & 60 \\
 \hline
 $3A$ &  $\sigma^a_i\sigma^a_j\sigma^a_k$&  $XXXII,YIYIY$ & 30 \\ 
 $3B$ & $\sigma^a_i\sigma^a_j\sigma^b_k$&  $IIXXY,ZXIXI,IYZIY$ & 180 \\ 
$3C$ & $\sigma^a_i\sigma^b_j\sigma^c_k$&  $IIXYZ,XZIYI,IYZIX$ & 60 \\
\hline
$4A$ & $\sigma^a_i\sigma^a_j\sigma^b_k\sigma^b_l$&  $IXXYY,XZXIZ,YZIZY$ & 90 \\
$4B$ & $\sigma^a_i\sigma^a_j\sigma^a_k\sigma^b_l$&  $XXXYI,XZZIZ,YZIYY$ & 120 \\
$4C$ & $\sigma^a_i\sigma^a_j\sigma^b_k\sigma^c_l$&  $XXZYI,YZZIX,YZIXY$ & 180 \\
$4D$ & $\sigma^a_i\sigma^a_j\sigma^a_k\sigma^a_l$&  $XXXXI,YIYYY$ & 15 \\
\hline
$5A$&$\sigma^a_i\sigma^a_j\sigma^a_k\sigma^a_l\sigma^b_m$&  $XXXXY,YZYYY$ & 30 \\
$5B$&$\sigma^a_i\sigma^a_j\sigma^a_k\sigma^b_l\sigma^b_m$&  $XYXXY,YZYZY$ & 60 \\
$5C$&$\sigma^a_i\sigma^a_j\sigma^b_k\sigma^b_l\sigma^c_m$&  $XYXZY,YYZZX$ & 90 \\
$5D$&$\sigma^a_i\sigma^a_j\sigma^a_k\sigma^b_l\sigma^c_m$&  $XXXZY,YZYXY$ & 60 \\
$5E$&$\sigma^a_i\sigma^a_j\sigma^a_k\sigma^a_l\sigma^a_m$&  $X^{\otimes 5},Y^{\otimes 5},Z^{\otimes 5}$ & 3 \\
\hline
\end{tabular}
\end{center}
\caption{Groupings of Pauli errors into sets of the same type and Hamming weight for the five-qubit code.}
\label{tab:5qubitgroups}
\end{table}

\begin{figure}[h!]
\centering
\includegraphics[width=0.45\textwidth]{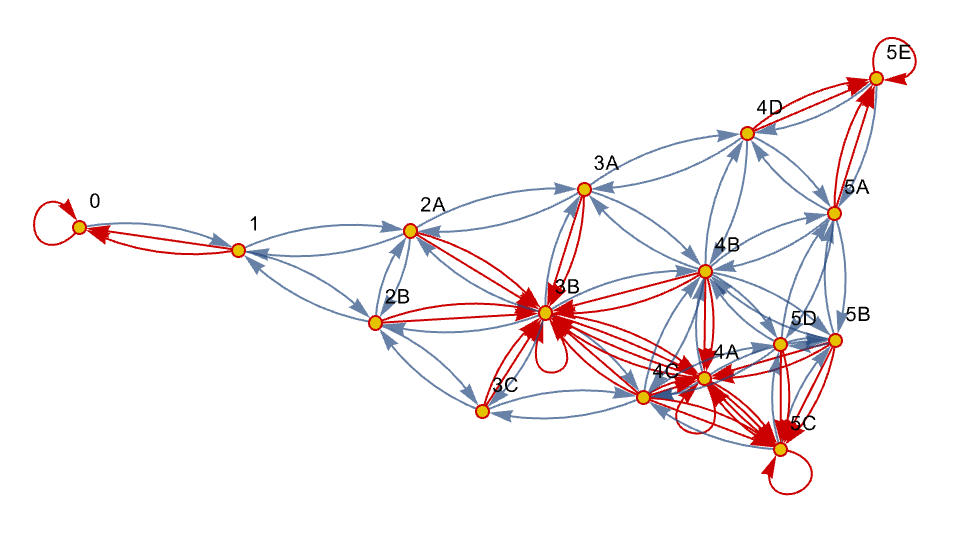}
\caption{Chain diagram for the five-qubit code with Markovian depolarizing noise and continuous error correction, showing the allowed  transitions among the coefficients from Pauli errors in blue and from the error-correcting process in red.}
\label{fig:chain5markov}
\end{figure}

We assume a depolarizing channel, where any weight-one Pauli error can occur with the same error rate $\gamma$:
\begin{eqnarray}\label{eq:Diss_five}
\gamma \mathcal{D}(\rho)&=&\gamma \sum_{i=1}^5 \sum_{a=x,y,z} (\sigma^a_i\rho \sigma^a_i -\rho)\\
&=& \gamma \sum_{i=1}^5 \left( X_i\rho X_i + Y_i\rho Y_i + Z_i\rho Z_i \right) - 15\gamma\rho . \nonumber
\end{eqnarray}

In Appendix~\ref{app:five_qubit_code}, we show the jump matrices in Eqs.~(\ref{eq:L1_5q}) and (\ref{eq:L0_5q}) that are associated with the error dissipator from Eq.~(\ref{eq:Diss_five}) and the error-correcting generator, respectively. Fig.~\ref{fig:chain5markov} shows the chain diagram for the five-qubit code with Markovian noise and continuous error correction.  The blue arrows show the transitions among the coefficients due to the depolarizing channel (\ref{eq:Diss_five}) and the red arrows show the error-correcting transitions. The exact jump rates are in the jump matrices.

\begin{figure}[h!]
	\centering
	\includegraphics[width=0.45\textwidth]{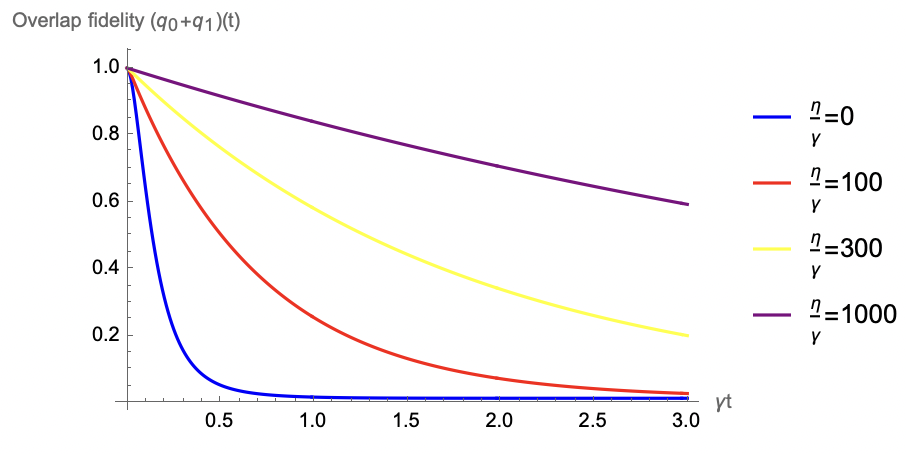}
	\caption{Overlap fidelity $q_0(t)+q_1(t)$ for the five-qubit code in the Markovian case as a function of dimensionless time $\gamma t$ for different values of the ratio ${\eta}/{\gamma}$.}
 \label{fig:fiveqmarkov}
\end{figure}

Similarly to the previous cases, in the long-time limit, if the error dissipation rate is much greater than the correction rate, the final state tends towards the maximally mixed state, where all $E_k\rho(0)E_k$ have the same weight.

By contrast, if the error-correction rate is much greater than the error rate the state will evolve to a mixture of the initial state with no errors $\rho_0$ and the state components with logical errors: $\rho_{5E}$, $\rho_{3B}$, $\rho_{4A}$, and $\rho_{5C}$.
In the long run for this case the density matrix converges to
\begin{equation}
\lim_{t \rightarrow \infty} \rho(t) = \rho_{0} + 30\rho_{3B} + 15\rho_{4A} + 15\rho_{5C} + \rho_{5E} .
\end{equation}

Fig.~\ref{fig:fiveqmarkov} shows the dynamics of the overlap fidelity $q_0(t)+q_1(t)$ for the five-qubit code in the Markovian case, The stronger the error-correction rate, the closer to fidelity one. The fidelity at short times is 
\begin{equation}
F_{M}(t) = 1 - 90\gamma^2 t^2 + 30 \gamma^2 (30\gamma+\eta) t^3 + O(t^4) .
\end{equation}
The leading order in times decays quadratically, similarly to the three-qubit case in Eq.~(\ref{markov3t2}).

\subsubsection{PMME for the five-qubit code}

\begin{figure}[h!]
	\centering
	\includegraphics[width=0.5\textwidth]{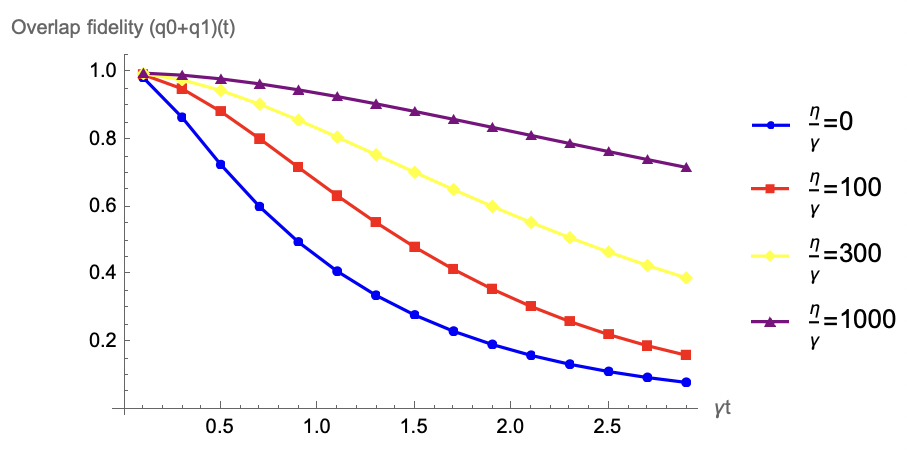}
	\caption{PMME overlap fidelity $q_0(t)+q_1(t)$ for the five qubit code with an exponential kernel $e^{-t}$ as a function of dimensionless time $\gamma t$ for different values of the ratio $\frac{\eta}{\gamma}$.}
 \label{fig:PMMEfiveq}
\end{figure}

Since the five-qubit code's superoperator involves a high-order characteristic polynomial, its eigenvalues cannot be expressed in closed analytical form. An analytical solution to the $X$-$X$ model is essentially out of reach. For this reason, we will only use the PMME model in Eq.~ (\ref{PMME}) to describe the five-qubit code with non-Markovian errors. We use the exponentially decaying kernel $ce^{-ct}$. As previously discussed, in the limit where $c\rightarrow \infty$ we recover the Markovian case. We will use the Laplace transform from Eq.~(\ref{eq:laplace_transf_PMME_3q}) with the matrix form of the dissipator from Eq.~(\ref{eq:L1_5q}) and the error correction from Eq.~(\ref{eq:L0_5q}).

In Fig.~\ref{fig:PMMEfiveq} we show the overlap fidelity for the PMME model with $c=1$. It shows a stark contrast with the Markovian case in Eq.~(\ref{markov3t2}): for the same error-correction rate $\eta$ the fidelity is higher in the non-Markovian model, due to the Zeno effect. In fact, up to leading order in time the fidelity goes as
\begin{equation}
F_{PMME}(t) = 1 - 30\gamma^3 t^3 + O(t^4) .
\end{equation}
Therefore, we conclude that continuous QEC performs better against non-Markovian noise and Markovian noise.

\section{Conclusion and Future Work}
\label{sec:conclusions}

In this paper, we investigated continuous quantum error correction (CQEC) and compared its performance with a Markovian error model against two different non-Markovian error models: an interaction Hamiltonian between the system and an environment qubit coupled to a ``cooling'' bath, and the post-Markovian master equation (PMME) with exponential bath memory kernels. In the Hamiltonian bath interaction, we observed abrupt transitions between Markovian and non-Markovian behavior and computed a measure of non-Markovianity based on the complete-positivity criterion. 

Our most important result is that, for short times, the fidelity in the non-Markovian models decays with leading order one power of $t$ greater than their Markovian counterpart, leading us to conclude that CQEC's performance is enhanced against non-Markovian interactions. This aligns with the quantum Zeno effect, where the state of the system ``freezes'' at short times, thereby protecting it from errors. Numerical results show that this advantage persists at longer times as well.

Finally, we suggest that this paper could serve as a tutorial for working with CQEC master equations analytically. It also lays the foundation for more complex codes; for starters, it should be easy to generalize it to the seven-qubit Steane code or the nine-qubit Bacon-Shor code. We expect a similar behavior where, at very long times, the state would evolve to be a mixture of logical errors, but at short times, the fidelity of the Markovian case would decay faster than the non-Markovian case. However, the method is not straightforwardly scalable to very large codes, as it would require measurements on high-weight operators in general. It could conceivably be applied to codes with low-weight stabilizer generators, such as the surface code. Another promising future direction is to model our approach using a stochastic master equation to include measurement currents and apply a feedback procedure as in \cite{Ahn2002,Oreshkov2007,Hsu2016}.

\acknowledgments

JGN thanks Professors Daniel A. Lidar and Benjamin W. Reichardt for inspiring work on this topic during their graduate courses. JGN and TAB also thank Zihan Xia, Maria Gabriela Boada Gutierrez, Anirudh Lanka, Dawei Zhong, Bruno Avritzer, Chris Gerhard, Onkar Apte, Yucheng He and Jagannathan Arjun Sathyamoorthy for fruitful discussions. This research was supported in part by NSF Grant No. 1911089 and by the U. S. Army Research Office under contract number W911NF2310255.

\appendix
\section{One qubit with continous error correction}

\subsection{Markovian bit flip dissipation}
\label{app:Markov_1q}

Using the Pauli basis $\mathcal{F}_P = \{I, X, Y, Z\}$, we can write the system described in Eq.~(\ref{simplest}) where the density matrix is represented as a $4\times 1$ vector ${v}_t^{S}$ that satisfies the linear differential equation 
\begin{equation}
\partial_t \mathbf{v}_t^{SB} = M	\mathbf{v}_0^{SB} ,
\end{equation}
with initial condition
\begin{equation}
\mathbf{v}_{0,P}^S = \frac{1}{2}(1,x_0,y_0,z_0) ,
\end{equation}
and the associated superoperator from Eq.~(\ref{simplest}) in matrix form is
\begin{equation}
M_P = \begin{pmatrix}
 0&0&0&0 \\ 0&-\eta&0&0 \\ 0&0&-(2\gamma+\eta)&0 \\
 \eta&0&0&-(2\gamma+\eta) \end{pmatrix} ,
\end{equation}
which has eigenvalues $\lambda=\{0,-(2\gamma+\eta),-(2\gamma+\eta),-\eta\}$. Therefore, the solution of the density matrix at time $t$ is
\[
\mathbf{v}_t^{SB}=\exp(Mt)\mathbf{v}_0^{SB} , 
\]
which is given in Eq.~(\ref{solsimplest}).

We can also use the computational basis $\mathcal{F}_C=\{\ket{0}\!\bra{0},\ket{0}\!\bra{1},\ket{1}\!\bra{1},\ket{0}\!\bra{1}\}$ to arrive at the same result. The corresponding initial state would be
\begin{equation}
	\mathbf{v}_{0,C}^S=(C_{0,0},C_{0,1},C_{1,0},C_{1,1})
\end{equation}
such that $C_{0,0}+C_{1,1}=1$ and the associated superoperator matrix $M$ is 
\begin{equation}
	M_C=\begin{pmatrix}
		-\gamma&0&0&\gamma+\eta\\0&-(\gamma+\eta)&\gamma&0\\0&\gamma&-(\gamma+\eta)&0\\ \gamma&0&0&-(\gamma+\eta)
	\end{pmatrix} ,
\end{equation}
which leads to the following vector at time $t$, related to the entries of the density matrix
\begin{equation}\label{ascend}
\begin{split}
\mathbf{v}_{t,C}^S=&\bigg(\frac{\gamma+\eta}{2\gamma+\eta}+\frac{e^{-(2\gamma +\eta)t}}{2\gamma+\eta}(\gamma C_{0,0}-(\gamma+\eta)C_{1,1}), \\
&e^{-(\gamma +\eta)t}(C_{0,1}\cosh \gamma t+C_{1,0}\sinh{\gamma t}),\\
&e^{-(\gamma +\eta)t}(C_{1,0}\cosh \gamma t+C_{0,1}\sinh \gamma t),\\
&\frac{\gamma}{2\gamma+\eta}
-\frac{e^{-(2\gamma +\eta)t}}{2\gamma+\eta}(\gamma C_{0,0}-(\gamma+\eta)C_{1,1})\bigg) .
\end{split}
\end{equation}
Again we obtain the limits for small and large correction coefficient
\begin{equation}
	\lim_{\gamma\rightarrow \infty}\mathbf{v}_t^S =
	\frac{1}{2} (1,0,0,1), \ \ \  \lim_{\eta\rightarrow \infty}\mathbf{v}_t^S=(1,0,0,0).
\label{eq:markovlimits}
\end{equation}

It is worth noticing that obtaining the superoperator matrix in the computational basis is straightforward, but it is easier to analyze the density matrix in the Pauli basis. We can easily jump from one representation to the other by a change of basis:
\begin{equation}\label{limitsbasura}
	\Lambda_{P\rightarrow C} = \begin{pmatrix}
		1&0&0&1 \\ 0&1&-i&0 \\ 0&1&i&0 \\1 &0&0&-1
	\end{pmatrix} ,
\end{equation}
so $\mathbf{v}_{0,C}^S=\Lambda_{P\rightarrow C}\mathbf{v}_{0,P}^S$ and $M_P=\Lambda_{P\rightarrow C}^{-1}M_C\Lambda_{P\rightarrow C}.$

An important feature about the linear differential matrix equations is that the solution state at long times converges to the stationary solution, that is simply the nullspace of the superoperator matrix:
\begin{eqnarray}
M_C\mathbf{v}_{\infty,C}=\mathbf{0} &\implies& \mathbf{v}_{\infty,C}=\frac{\gamma+\eta}{2\gamma+\eta} \left(1,0,0,\frac{\gamma}{\gamma+\eta}\right) , \nonumber\\
M_P\mathbf{v}_{\infty,P} = \mathbf{0} &\implies& \mathbf{v}_{\infty,P} = \frac{1}{2} \left(1,0,0,\frac{\eta}{2\gamma+\eta}\right) .
\end{eqnarray}
Then the stationary state at long times is 
\begin{eqnarray}
\lim_{t\rightarrow \infty}\rho_t^S &=& \frac{1}{2} \left(I^S+\frac{\eta}{2\gamma+\eta}\sigma_z^S\right) \\
&=& \frac{\gamma+\eta}{2\gamma+\eta} \ket{0}^S\!\bra{0}^S + \frac{\gamma}{2\gamma+\eta}\ket{1}^S\!\bra{1}^S, \nonumber
\end{eqnarray}
which reduces to the limits given previously in Eq.~(\ref{eq:markovlimits}) when the correction or dissipation coefficients become large.

\subsection{System-bath $X$-$X$ coupling with a cooling bath}\label{app:XX_1q}

To solve the coupled system in Eq.~(\ref{couplingbathgamma}) with an initial system state given by Eq.~(\ref{init}) we use the operator space of the system and bath with the basis $\{\sigma_i^S\otimes \sigma_j^B\}$ (where $\sigma_{0,1,2,3} \equiv I,X,Y,Z$). The superoperator can then be written as a $16\times 16$ matrix and the joint system-bath density matrix as a $16\times 1$ vector.
\begin{equation}\label{inivector}
	\mathbf{v}_0^{SB}=\frac{1}{4}(1,0,0,1,x_0,0,0,x_0,y_0,0,0,y_0,z_0,0,0,z_0)^T.
\end{equation}
The master equation Eq.~(\ref{couplingbathgamma}) can be written in matrix form as
\[
\partial_t 	\mathbf{v}_t^{SB} = M\mathbf{v}_0^{SB} ,
\]
where the matrix $M$ is given by
\begin{widetext}
\begin{equation}	\hspace{-0.5cm}M\hspace{-0.1cm}=\hspace{-0.1cm}\begin{pmatrix}
0& 0& 0& 0& 0& 0& 0& 0& 0& 0& 0& 0& 0& 0& 0& 0\\
0& -\frac{\kappa}{2}& 0& 0& 0& 0& 0& 0& 0& 0& 0& 0& 0& 0& 0& 0\\
0& 0&-\frac{\kappa}{2}& 0& 0& 0& 0& -2\alpha& 0& 0& 0& 0& 0& 0& 0& 0\\
\kappa& 0& 0& -\kappa& 0& 0& 2\alpha& 0& 0& 0& 0& 0& 0& 0& 0& 0\\
0& 0& 0& 0& -\eta& 0& 0& 0& 0& 0& 0& 0& 0& 0& 0& 0\\
0& 0& 0& 0& 0& -\eta-\frac{\kappa}{2}& 0& 0& 0& 0& 0& 0& 0& 0& 0& 0\\
0& 0& 0& -2\alpha& 0&0& -\eta-\frac{\kappa}{2}& 0& 0& 0& 0& 0& 0& 0& 0& 0\\
0& 0& 2 \alpha& 0& \kappa& 0& 0& - \eta-\kappa& 0& 0& 0& 0& 0&0& 0& 0&\\
0& 0& 0& 0& 0& 0& 0& 0& -\eta& 0& 0& 0& 0& -2 \alpha& 0& 0\\ 
		0& 0& 0& 0& 0& 0& 0& 0& 0& - \eta-\frac{\kappa}{2}& 0& 0&-2 \alpha& 0& 0& 0\\
		0& 0& 0& 0& 0& 0& 0& 0& 0& 0&-\eta-\frac{\kappa}{2}& 0& 0& 0& 0& 0\\
		0& 0& 0& 0& 0& 0& 0& 0& \kappa& 0& 0& - \eta-\kappa& 0& 0& 0& 0\\
		\eta& 0& 0& 0& 0& 0&0& 0& 0& 2 \alpha& 0& 0& -\eta& 0& 0& 0\\ 
		0&\eta& 0& 0& 0& 0& 0& 0& 2 \alpha& 0& 0& 0& 0& -\eta-\frac{\kappa}{2}& 0& 0\\ 0&0&\eta& 0& 0& 0& 0& 0& 0& 0& 0& 0& 0& 0& - \eta -\frac{\kappa}{2}& 0\\
		0& 0& 0& \eta& 0& 0& 0& 0& 0& 0& 0& 0& \kappa& 0& 0& -\eta-\kappa\\
	\end{pmatrix}.
	\end{equation}
\end{widetext}

The solution to $\mathbf{v}_t^{SB}$ satisfies 
\[
\mathbf{v}_t^{SB} = \exp(Mt)\mathbf{v}_0^{SB}
\]
and from it we can build the system state described in Eq.~(\ref{evolr}), where the time-dependence of the coefficients is given in Eqs.~(\ref{eq:c_coeff_XX}) and (\ref{eq:d_coeff_XX}).

\subsubsection{Trace distance measure}

Let us analyze the Markovian behavior using the measure $\mathcal{N}$ from Eq.~(\ref{measure}) and the expression for the density matrix at time $t$ from Eq.~(\ref{evolr}):
\begin{equation}
\rho_t^S-{\rho'}_t^S = \frac{1}{2} \left[e^{-\eta t} \Delta_x X^S + \mathfrak{C}(t)(\Delta_y Y^S + \Delta_z Z^S)\right] .
\end{equation} 
The trace distance is 
\begin{equation}
d_{\mathrm{Tr}} (\rho_t^S,{\rho'}_t^S) = \frac{1}{2} \left[ e^{-2\eta t}\Delta_x^2 + (\mathfrak{C}(t))^2 (\Delta_y^2 + \Delta_z^2) \right]^{1/2} ,
\end{equation} 
and its time derivative is
\begin{eqnarray}
\partial_t d_{Tr}(\rho_t^S,{\rho'}_t^S) &=& 
\left(- \frac{\eta e^{-2\eta t}}{\mathfrak{C}^2(t)} \Delta_x^2 + \frac{\mathfrak{C}'(t)} {\mathfrak{C}(t)} (\Delta_y^2+\Delta_z^2)\right) \nonumber\\
&& \times \frac{\mathfrak{C}^2(t)}{2d_{\mathrm{Tr}}(\rho_t^S,{\rho'}_t^S)} .
\end{eqnarray}

If we consider initial states with the same $X$ coefficient $x_0$, in other words $\Delta_x=0$, the trace distance increases when 
\begin{equation}
0 < \frac{\mathfrak{C}'(t)}{\mathfrak{C}(t)} = -\frac{\eta \kappa+16\alpha^2} {\kappa+\sqrt{64\alpha^2-\kappa^2} \cot \frac{t}{4} \sqrt{64\alpha^2-\kappa^2}} ,
\end{equation}
That is equivalent to the inequality
\[
\cot \frac{t}{4}\sqrt{64\alpha^2-\kappa^2} < -\frac{\kappa} {\sqrt{64\alpha^2-\kappa^2}}. 
\]
The time intervals when this occurs are $t_n < t < t_n-\delta$ for positive integers $n$, with
\begin{equation}
t_n = \frac{4n\pi}{\sqrt{64\alpha^2-\kappa^2}} \ \ \  \delta = \frac{\arctan\left(\frac{1}{\kappa} \sqrt{64\alpha^2 - \kappa^2}\right)}{\sqrt{64\alpha^2-\kappa^2}} ,
\end{equation}
where
\[
\mathfrak{C}(t_n) = e^{-(\eta + \kappa/4) t_n} \cos(n\pi) = (-1)^ne^{-(\eta+\frac{\kappa}{4})t_n} ,
\]
and $\mathfrak{C}(t_n-\delta)=0$.

For $\kappa^2 < 64\alpha^2$, we can find the maximum by using the triangle inequality over two vectors $\mathbf{x}_0$ and $\mathbf{x}'_0$ with norm at most 1, such that $\sqrt{\Delta_y^2+\Delta_x^2}=||\mathbf{x}_0-\mathbf{x}'_0||\leq 2$:
\begin{equation}
\begin{split}
\max d_{\mathrm{Tr}}(\rho_t^S,{\rho'}_t^S) &= \max_{\Delta_y,\Delta_z} \frac{1}{2} \left[(\mathfrak{C}(t_n))^2 (\Delta_y^2+\Delta_z^2) \right]^{1/2} \\
&= e^{-(\eta+\frac{\kappa}{4})t_n} .
\end{split}
\end{equation}

Therefore, summing this last quantity over all integers we get the non-Markovianity measure in Eq.~(\ref{Nonmarknew}).
For nonzero dissipation and correction rates the result for the measure is a finite number, a weaker non-Markovian behavior. For no dissipation, $\kappa=0$, the result is 
\begin{equation}
\mathcal{N} = \frac{1}{e^{\frac{\eta \pi}{2\alpha}}-1} ,\ \ \ \kappa=0.
\end{equation}
For a higher correction rate, $\eta \gg \kappa$, we recover the Markovian case,
\[
\mathcal{N} \approx e^{-\frac{\eta \pi}{2\alpha}} \rightarrow 0 .
\]
This is quite reasonable, since a stronger correction reduces the system qubit to the desired state $\ket{0}_S \bra{0}_S$ regardless of the initial state.

\subsection{One qubit Post-Markovian Master Equation}
\label{app:PMME_1q}

We can express the one-qubit state in the basis of Pauli matrices. Assuming again the initial state from Eq.~(\ref{init}), the state at time $t$ is given by Eq.~(\ref{PMMEsOL}). We now derive it. Beginning with 
\begin{equation}
\rho(t) = a(t)I + x(t)X + y(t)Y + z(t)Z .
\end{equation}
Using the PMME equation (\ref{PMME}), we arrive at the following differential equations for $a(t)$, $x(t)$, $y(t)$ and $z(t)$:
\begin{equation}
\begin{split}
	\dot{a}(t) =&\ 0 , \\
	\dot{x}(t) =& -\eta x(t) , \\
	\dot{y}(t )=& -\eta y(t) + \int_{0}^t k(t') \left( -2\gamma e^{-(2\gamma + \eta)t'}\right) y(t-t') dt' , \\
	\dot{z}(t) =& -\eta a(t)-\eta z(t) \\
    &+ \int_{0}^t k(t') \biggl(-\frac{2\gamma \eta}{2\gamma+\eta}  (1-e^{-(2\gamma+\eta)t'})a(t-t') \\
    &\qquad - 2\gamma e^{-(2\gamma+\eta)t'}z(t-t')\biggr) dt' .
\end{split}
\end{equation}
The solutions to the first two equations are immediate:
\begin{equation}
		a(t)=\frac{1}{2}, \, \ \ \ 
		x(t)=\frac{1}{2} x_0e^{-\eta t}.
\end{equation}
The last two integro-differential equations depend on the kernel, and the Laplace transform method will be used to solve them. We apply the Laplacian transformation to the differential equation for $y(t)$:
\begin{equation}
\begin{split}
s\tilde{y}(s) - y(0) =& -\eta \tilde{y}(s) + (-2\gamma) Lap[k(t)e^{-(2\gamma+\eta)t}] \tilde{y}(s) \\
=& -\eta \tilde{y}(s) + (-2\gamma) \tilde{k}(s+2\gamma+\eta) \tilde{y}(s) ,
\end{split}
\end{equation}
where the tilde denotes the Laplace transform of the variable. Then,
\[
y(t)=\frac{y_0}{2}\xi(t) ,
\]
where
\begin{equation}\label{xit}
\xi(t) = Lap^{-1}\left(\frac{1}{s+\eta+2\gamma \tilde{k}(s+2\gamma+\eta)}\right).
\end{equation}

For $z(t)$ we can also use the Laplace transform:
\begin{equation}
\begin{split}
	s\tilde{z}(s)-z(0) =&\ \eta\tilde{a}(s) - \eta\tilde{z}(s) - \frac{2\gamma \eta}{2\gamma+\eta} \tilde{k}(s) \tilde{a}(s) \\
    &+ \frac{2\gamma \eta}{2\gamma+\eta} 
	\tilde{k}(s+2\gamma+\eta) \tilde{a}(s) \\
    &- 2\gamma\hat{k}(s+2\gamma+\eta) \tilde{z}(s) ,
\end{split}
\end{equation}
which gives a solution
\begin{equation}
z(t) = \frac{z_0}{2}\xi(t)+ \chi(t) .
\end{equation}
Using $\tilde{a}(s)=1/2s$, $\chi(t)$ is 
\begin{equation}
\begin{split}\label{chit}
\chi(t) =&\ \eta Lap^{-1}\left[\frac{\frac{1}{2s} }{s+\eta+2\gamma \hat{k}(s+\eta+2\gamma)}\right] \\
	&+ \frac{\gamma \eta}{2\gamma+\eta} Lap^{-1}\left[\frac{ (\tilde{k}(s + 2\gamma + \eta) - \tilde{k}(s))/2s }{s + \eta + 2\gamma \hat{k}(s + \eta + 2\gamma)}\right].
\end{split}
\end{equation}

\subsubsection{Decaying exponential kernel}

For the exponential kernel $k = ae^{-ct}$,  with Laplace transform $\tilde{k} = a/(s+c)$, and assuming that $\left(\frac{c}{2}+\gamma\right)^2 > 2\gamma a$, the state at time $t$ is
\begin{equation}
\rho(t) = \frac{1}{2} (I + x_0 e^{-\eta t} X + \xi(t)(y_0 Y + z_0 Z)) + \chi(t)Z ,
\end{equation}
where
\begin{equation}
\begin{split}
\xi(t) =&\ e^{-(\frac{c}{2}+\gamma+\eta)t} \Biggl(\frac{(\frac{c}{2}+\gamma) \sinh t\sqrt{\left(\frac{c}{2}+\gamma\right)^2-2\gamma a}}{\sqrt{\left(\frac{c}{2}+\gamma\right)^2-2\gamma a}} \\
& + \cosh t\sqrt{\left(\frac{c}{2}+\gamma\right)^2-2\gamma a} \Biggr) ,
\end{split}
\end{equation}

\begin{equation}
\begin{split}
\chi(t) =&\, \frac{\eta}{2} \cdot \frac{c + 2\gamma + \eta}{2\gamma a + \eta(c + 2\gamma + \eta)} \\
&+ \frac{\eta}{2} e^{-\left(\frac{c}{2} + \gamma + \eta\right)t} \Biggl[ 
\left( \frac{4a\gamma - (c + 2\gamma)(c + 2\gamma + \eta)}{2a\gamma + \eta(c + 2\gamma + \eta)} \right) \\
&\quad\times \frac{\sinh\left( \frac{t}{2} \sqrt{(c + 2\gamma)^2 - 8a\gamma} \right)}{\sqrt{(c + 2\gamma)^2 - 8a\gamma}} \\
&- \frac{c + 2\gamma + \eta}{2a\gamma + \eta(c + 2\gamma + \eta)} \\
&\quad\times \cosh\left( \frac{t}{2} \sqrt{(c + 2\gamma)^2 - 8a\gamma} \right)
\Biggr] \\
&+ \frac{a\gamma\eta}{-2\gamma a + (c - \eta)(2\gamma + \eta)} \Biggl[
e^{-ct} - e^{-\left( \frac{c}{2} + \gamma + \eta \right)t} \\
&\quad\times \biggl(
\cosh\left( t \sqrt{\left( \frac{c}{2} + \gamma \right)^2 - 2\gamma a} \right) \\
&- \frac{(c - 2(\gamma + \eta)) \sinh\left( t \sqrt{\left( \frac{c}{2} + \gamma \right)^2 - 2\gamma a} \right)}{2 \sqrt{\left( \frac{c}{2} + \gamma \right)^2 - 2\gamma a}}
\biggr)
\Biggr]
\end{split}
\end{equation}

\subsubsection{Decaying oscillating damped exponential kernel}

For the overdamped or underdamped situation of Eq.~(\ref{kernel2}) we note that this kernel will cause that the time-dependent coefficients $\xi(t)$ and $\chi(t)$ to have a cubic polynomial denominator whose roots, unfortunately, cannot be expressed as a simple algebraic solutions. For example,
\begin{eqnarray}
\xi(t) &=& Lap^{-1}\left(\frac{1}{s + \eta + 2\gamma \tilde{k}(s + 2\gamma + \eta)}\right) \\
&=& Lap^{-1}\left(\frac{1}{s + \eta + 2\gamma \frac{s + \eta + 2\gamma + a}{(s + \eta + 2\gamma)^2 + b (s + \eta + 2\gamma) + c} }\right) \nonumber
\end{eqnarray}
will include an exponential of $t$ multiplying the roots of the polynomial 
\begin{equation}
\begin{split}
p_3(x) =&\ 4 + 2 a + 6 r + 2 b r + c r + 4 r^2 + b r^2 + r^3 \\
&+ (6 + 2 b + c + 8 r/+ 2 b r + 3 r^2) x \\
&+ (4 + b + 3 r) x^2 + x^3 = 0 ,
\end{split}
\end{equation}
where $a$, $b$ and $c$ have the same units as $\gamma$ (inverse time) and $r=\eta/\gamma$.

\section{Fidelity and the Zeno Effect}
\label{app:fidelity}

\subsection{Non-Markovian system-bath coupling with error correction and cooling bath}

Combining the expression for the fidelity from Eq.~(\ref{fidelity}) with the solution of the non-Markovian system from Eq.~(\ref{evolr}), with system-bath $X$-$X$ coupling at rate $\alpha$, a cooling bath at rate $\kappa$, and error correction at rate $\eta$, yields two cases.


For the non-Markovian case $\kappa^2<64\alpha^2$
\begin{equation} \label{nonmarkfidel}
\begin{split}
F_{XX,NM}(t) =& \frac{\eta(\kappa +2\eta)+4\alpha^2}{\eta(\kappa +2\eta)+8\alpha^2} \\
&+ \frac{4\alpha^2}{\eta(\kappa +2\eta)+8\alpha^2} e^{-(\eta+\frac{\kappa}{4})t} \\
&\times\biggl[(4\eta+\kappa) \frac{\sin (t/4) \sqrt{64\alpha^2-\kappa^2} }{\sqrt{64\alpha^2-\kappa^2}} \\
&\qquad +\cos (t/4) \sqrt{64\alpha^2-\kappa^2} \biggr] .
\end{split}
\end{equation}
For no dissipation from the cooling bath ($\kappa=0$)  and only error correction we get the fidelity term obtained by [4] Oreshkov.

For the Markovian case $\kappa^2> 64\alpha^2$
\begin{equation}
\begin{split} \label{markfidel}
F_{XX,M}(t) =& \frac{\eta(\kappa +2\eta)+4\alpha^2}{\eta(\kappa +2\eta)+8\alpha^2} \\
&+ \frac{4\alpha^2}{\eta(\kappa +2\eta)+8\alpha^2} e^{-(\eta+\frac{\kappa}{4})} \\
&\times \biggl[(4\eta+\kappa)\frac{ \sinh \frac{t}{4}\sqrt{\kappa^2-64\alpha^2} }{\sqrt{\kappa^2-64\alpha^2}} \\
& \qquad +\cosh \frac{t}{4} \sqrt{\kappa^2-64\alpha^2}\biggr].
\end{split}
\end{equation}

For the Markovian case $\kappa^2 = 64\alpha^2$ (or $\kappa = 8\alpha$),
\begin{equation}
\begin{split}
F_{XX,M}(t) =& \frac{1}{2}\biggl(1+e^{-\left(\eta+2\alpha\right)t}\left(1+2\alpha t\right) \\
&+ \frac{\eta(\eta+4\alpha)}{(\eta+2\alpha)^2} 
\biggl(1 + e^{-\left(\eta+2\alpha\right)t} \\
& \quad \times\left(\left(\frac{4\alpha^2}{4\alpha+\eta}-2\alpha\right)t-1\right)\biggr)\biggr) .
\end{split}
\end{equation}

\subsection{The three-qubit code}
\label{app:3qcode}

\begin{figure}[h!]
\centering
\begin{quantikz}
\lstick{$\ket{\psi}$} & \ctrl{1} & \ctrl{2} & \qw \\
\lstick{$\ket{0}$} & \targ{} & \qw & \qw \\
\lstick{$\ket{0}$} & \qw & \targ{} & \qw
\end{quantikz}
\caption{Encoding circuit for the three-qubit bit-flip code. 
}
\label{3qcircuit}
\end{figure}
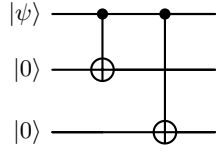

To encode a single qubit $a\ket{0}+b\ket{1}$ into the three-qubit state  $a\ket{000}+b\ket{111}$ we can construct \cite{Nielsen} a simple encoding unitary:
\begin{equation}
U_{\mathcal{E}} = \ket{0}\bra{0}\otimes I\otimes I + \ket{1}\bra{1}\otimes X\otimes X,
\end{equation}
which can by simply constructed with a quantum circuit as shown in Fig.~\ref{3qcircuit}.

For the bit-flip channel there are four error syndromes corresponding to the four projection operators
\begin{equation}
\begin{split}
P_0\equiv \ket{000}\bra{000}+ \ket{111}\bra{111}, \\
P_1\equiv \ket{100}\bra{100}+ \ket{011}\bra{011}, \\
P_2\equiv \ket{010}\bra{010}+ \ket{101}\bra{101}, \\
P_3\equiv \ket{001}\bra{001}+ \ket{110}\bra{110},
\end{split}
\end{equation}
which help us identify (detect) the corresponding error. For instance, the ``erroneous'' state $a\ket{010}+b\ket{101}$, with an error in the second qubit, can be detected with the projector $P_2$. Note that applying this projector will not change the state. To do the appropriate corrections, we need the following correcting unitaries:
\begin{equation}
\begin{split}
C_0 \equiv I , &\ \  C_1 \equiv X_1 , \\
C_2 \equiv X_2 , &\ \  C_3 \equiv X_3 .
\end{split}
\end{equation}
Hence the error correcting CPTP map \cite{Oreshkov2007} given in Eq.~(\ref{eq:CPTP_error}) can be written
\begin{equation}
	\begin{split}
		\Phi(\rho) =&\ 
(\ket{000}\!\bra{000}+\ket{111}\!\bra{111}) \\
& \times (\rho + X_1\rho X_1 + X_2\rho X_2 + X_3\rho X_3) \\
& \times (\ket{000}\!\bra{000}+\ket{111}\!\bra{111}) ,
\end{split}
\end{equation}
where we see that we can write the correction map as applying the error gates that could be corrected,
\[
\sum_{i=0}^3X_i\rho X_i ,
\]
and then applying the code space projector $\ket{000}+\ket{111}$. The error-correcting generator is $\Gamma(\rho) = \Phi(\rho) - \rho$ that is used in Eq.~(\ref{eq:3bit_error_gen}).

In this section, we will apply this error-correcting generator to the generalized three-qubit version of the models given for the one-qubit system.

\subsubsection{Markovian three-qubit code}

To solve the Markovian differential equation
\[
\partial_t \rho^S = \gamma \mathcal{D}(\rho) + \eta \Gamma(\rho) , 
\]
we can construct a vector 
\begin{equation}
\mathbf{v}_t^S = \left(\begin{array}{c} q_0(t) \\ q_1(t) \\ q_2(t) \\ q_3(t)) \end{array}\right) ,
\label{eq:rho3vector}
\end{equation}
with the coefficients of the proposed solution in Eq.~(\ref{rho3qubit}) and the matrix $M$ given by
\begin{equation}
M = \begin{pmatrix}
	-3\gamma & \eta + \gamma & 0 & 0 \\
    3\gamma & -(\eta+3\gamma) & 2\gamma & 0 \\
    0 & 2\gamma & -(\eta+3\gamma) & 3\gamma \\
    0 & 0 & \eta + \gamma & -3\gamma
\end{pmatrix} ,
\end{equation}
where $M$ is the jump matrix, where the columns sum up to $0$, and whose eigenvalues are
\begin{eqnarray*}
\lambda_0 &=& 0 , \\
\lambda_1 &=& -4\gamma - \eta ,\\
\lambda_{2,3} &=& \frac{1}{2}(-8\gamma - \eta \pm \sqrt{16\gamma^2 + 16\gamma+\eta^2}).
\end{eqnarray*}

Given an initial state $\mathbf{v}_0^S=(1,0,0,0)$ with $q_0(0)=1$, $q_1(0)=q_2(0)=q_3(0)=0$, the solution for any time $t$ is
\begin{equation}\label{3markov}
\begin{split}
	q_{0,3}(t) =& \frac{\eta + \gamma + 3\gamma e^{-(\eta+4\gamma)t}}{2(\eta+4\gamma)} \\
&\pm \frac{1}{2}e^{-(\frac{\eta}{2}+4\gamma)t} 
\biggl(\cosh \frac{t}{2}\sqrt{\eta^2+16\eta \gamma+16\gamma^2} \\
&\quad + \frac{(\eta+2\gamma) \sinh \frac{t}{2}\sqrt{\eta^2+16\eta \gamma+16\gamma^2} }{\sqrt{\eta^2+16\eta \gamma+16\gamma^2}} \biggr) , \\
q_{1,2}(t) =& \frac{3\gamma}{2} \biggl(\frac{1-e^{-(\eta+4\gamma)t}}{\eta+4\gamma}  \\
& \pm 2e^{-\left(\frac{\eta}{2}-4\gamma\right)t} 
\frac{\sinh \frac{t}{2}\sqrt{\eta^2+16\eta \gamma+16\gamma^2} }{\sqrt{\eta^2+16\eta \gamma+16\gamma^2}}\biggr) .
\end{split}
\end{equation} 
These coefficients were obtained by Ahn \cite{Ahn2002} in the case of no error correction.

In the long-time limit we get a stationary state:
\begin{equation}
\lim_{t\rightarrow \infty} \rho(t)^S=\frac{\eta+\gamma }{2(\eta+4\gamma)}\left(\rho(0)+\rho_3\right)+\frac{3}{2} \frac{\gamma}{\eta+4\gamma}(\rho_1+\rho_2) .
\end{equation}
According to Ahn, the recovered state, averaged over all possible measurement syndromes, is
\begin{equation}
\begin{split}
\rho(t)=&(q_0(t)+q_1(t))\rho(0)+\\&(q_2(t)+q_3(t))X_1X_2X_3\rho(0)X_1X_2X_3
\end{split}
\end{equation} 
The overlap of this state with the initial state depends on the initial state, if this is $\rho(0)=\ket{000}\bra{000}$ the logical zero of the code we get a fidelity $F(t)=q_0(t)+q_1(t)$:
\begin{equation}\label{markov3qfid}
\begin{split}
F(t) =& \frac{1}{2} + \frac{e^{-t(4\gamma+\frac{\eta}{2})}}{2} \Biggl(\cosh\left(\frac{t}{2} \sqrt{16\gamma^2+16\gamma\eta+\eta^2}\right) \\ 
&+ (8\gamma+\eta) \frac{\sinh\left(\frac{t}{2} \sqrt{16\gamma^2+16\gamma\eta+\eta^2} \right)}{\sqrt{16\gamma^2+16\gamma\eta+\eta^2}} \Biggr) ,
	\end{split}
\end{equation}
which converges to $1/2$ at long times. For very short times we can Taylor expand the matrix exponential up to quadratic order in time, as given by Eq.~(\ref{markov3t2}).

\subsubsection{Three-qubit PMME with exponential kernel}

The generators $\mathcal{L}_0$ and $\mathcal{L}_1$ can be expressed in matrix form over the basis $\{\rho(0),\rho_1,\rho_2,\rho_3\}$ whose linear combination gives the system state in Eq.~(\ref{rho3qubit}) as the vector $\mathbf{v}_t^S$ in Eq.~(\ref{eq:rho3vector}):
\begin{eqnarray}
\mathcal{L}_0 &=& \begin{pmatrix}
	0 & \eta & 0 & 0 \\
    0 & -\eta & 0 & 0 \\
    0 & 0 & -\eta & 0 \\
    0 & 0 & \eta & 0
\end{pmatrix} , \nonumber\\
\mathcal{L}_1 &=& \begin{pmatrix}	
-3\gamma & \gamma & 0 & 0 \\
3\gamma & -3\gamma & 2\gamma & 0 \\
0 & 2\gamma & -3\gamma & 3\gamma \\
0 & 0 & \gamma & -3\gamma
\end{pmatrix} ,
\end{eqnarray}
The eigenvalues of the matrix $[\mathcal{L}_0+\mathcal{L}_1]$ are
\begin{eqnarray*}
\lambda_0 &=&0 , \nonumber\\
\lambda_1 &=& -4\gamma-\eta \\
\lambda_{2,3} &=& \frac{1}{2}(-8\gamma - \eta \mp \sqrt{16\gamma^2+16\gamma+\eta^2}).
\end{eqnarray*}
The expression $\exp[(\mathcal{L}_0+\mathcal{L}_1)t']$ can be written as
\[
\exp[(\mathcal{L}_0 + \mathcal{L}_1)t'] = Q \Lambda(t) Q^{-1} ,
\]
where 
\begin{equation}
\Lambda(t) = \begin{pmatrix} 
e^{\lambda_0 t} & 0 & 0 & 0 \\
0 & e^{\lambda_1 t} & 0 & 0 \\
0 & 0 & e^{\lambda_2 t} & 0 \\
0 & 0 & 0 & e^{\lambda_3 t} \end{pmatrix} ,
\end{equation}
and $Q$ is the matrix whose columns are the eigenvectors of $(\mathcal{L}_0 + \mathcal{L}_1)$. The PMME equation for the coefficients $q_i(t)$ of Eq.~(\ref{rho3qubit}) can be written
\begin{equation}\label{eq:PMME_amaia}
\frac{d\mathbf{q}}{dt}(t) = \mathcal{L}_0\mathbf{q}(t) + \mathcal{L}_1 Q\int_{0}^{t} dt'\, k(t') \Lambda(t)  Q^{-1} \mathbf{q}(t-t').
\end{equation}
This equation can be solved using the Laplace transformation. Let $\tilde{\mathbf{q}}(s)$ be the Laplace transform of the 4-vector $\vec{q}(t)$, and $\tilde{k}_i = \tilde{k}(s-\lambda_i)$ be the Laplace transform of the kernel shifted by the correspondent eigenvalue $\lambda_i$ of $(\mathcal{L}_0+\mathcal{L}_1)$. Then Eq.~(\ref{eq:PMME_amaia}) transforms to
\begin{equation}\label{eq:laplace_transf_PMME_3q}
s\tilde{\mathbf{q}}(s) - \mathbf{q}(0) = \mathcal{L}_0 \tilde{\mathbf{q}}(s) + \mathcal{L}_1 Q\, \text{diag}[\tilde{k}_i]Q^{-1}\tilde{\mathbf{q}}(s) ,
\end{equation}
where $\text{diag}[\tilde{k}_i]$ indicates a diagonal matrix whose diagonal elements are the Laplace transforms $\tilde{k}_i$ defined above. We can solve for $\tilde{\mathbf{q}}(s)$, and then take the inverse Laplace transform of the entries to obtain $\mathbf{q}(t)$.

In the case of an exponential kernel $k(t)=ce^{-ct}$, its Laplace transform is
\[
\tilde{k}(s) = \frac{c}{s+c} .
\]
Hence, the overlap fidelity $q_0(t)+q_1(t)$ is
\begin{widetext}
\begin{eqnarray}\nonumber \label{fidpmme3qexp}
F(t) &=& q_0(t) + q_1(t) = \frac{1}{2} + \frac{1}{2}\frac{1}{1-\frac{8\gamma+\eta}{c} + 12\left(\frac{\gamma}{c}\right)^2} \Biggl(12\left(\frac{\gamma}{c}\right)^2 e^{-ct} + e^{-t(4\gamma+\frac{\eta}{2})} \biggl(\left(1 - \frac{8\gamma+\eta}{c} \right) \cosh\left(\frac{t}{2} \sqrt{16\gamma^2+16\gamma\eta+\eta^2}\right) \nonumber\\
&&+ \left(8\gamma\left(1-5\frac{\gamma}{c}\right) + \eta\left(1-16\frac{\gamma}{c}\right) - \frac{\eta^2}{c}\right) \frac{\sinh\left(\frac{t}{2}\sqrt{16\gamma^2+16\gamma\eta+\eta^2}\right)}{\sqrt{16\gamma^2+16\gamma\eta+\eta^2}}
\biggr)\Biggr).
\end{eqnarray}
\end{widetext}
At short times the expansion of the fidelity is given by Eq.~(\ref{fidpmme3q}).

\subsubsection{Three system and three bath qubits with $X$-$X$ coupling and cooling bath}

Briefly we describe the program to create the matrix in the computational basis for codes with large number of physical qubits. For instance, denoting the three system qubits with italic letters and the three environment qubits with greek letters, one component of the density matrix can be written
\begin{equation}
\begin{split}
\rho_x =& \ket{a_1 a_2 a_3 \alpha_1 \alpha_2 \alpha_3}\bra{b_1 b_2 b_3 \beta_1\beta_2\beta_3} \\
\equiv & \ket{a_1}\bra{b_1}  \otimes  \ket{a_2}\bra{b_2} \otimes \ket{a_3}\bra{b_3} \\
& \otimes  \ket{\alpha_1}\bra{\beta_1} \otimes \ket{\alpha_2}\bra{\beta_2} \otimes \ket{\alpha_3}\bra{\beta_3} ,
\end{split}
\end{equation}
to be written as a binary string
\[
x = a_1 a_2 a_3 \alpha_1 \alpha_2 \alpha_3 b_1 b_2 b_3 \beta_1 \beta_2 \beta_3 .
\]
The Hamiltonian in Eq.~(\ref{eq:XX_3qubit}) corresponds to terms like
\begin{equation}
\begin{split}
& a_1+1, a_2, a_3, \alpha_1+1,\alpha_2, \alpha_3, b_1, b_2, b_3, \beta_1, \beta_2, \beta_3 ,\\
&a_1, a_2, a_3, \alpha_1, \alpha_2, \alpha_3, b_1+1, b_2, b_3, \beta_1+1, \beta_2, \beta_3 , 
\end{split}
\end{equation}
and so forth. Similarly, one of the dissipator terms in Eq.~(\ref{eq:dissip3qubit}) acts as
\begin{equation}
\begin{split}
\mathcal{D}(\sigma_{1-})\rho_x =&\, \delta_{\alpha_1,1}\delta_{\beta_1,1}\ket{a_1 a_2 a_3 0 \alpha_2 \alpha_3}\bra{b_1 b_2 b_3 0 \beta_2 \beta_3}\\
& -\frac{1}{2}\delta_{\alpha_1,1}\ket{a_1 a_2 a_3 1 \alpha_2 \alpha_3}\bra{b_1 b_2 b_3 \beta_1 \beta_2 \beta_3} \\
& -\frac{1}{2}\delta_{\beta_1,1}\ket{a_1 a_2 a_3 \beta_1 \alpha_2 \alpha_3}\bra{b_1 b_2 b_3 1 \beta_2 \beta_3} .
\end{split}
\end{equation}
Finally, we can calculate the fidelity of the reduced density matrix:
\[
F = \mathbf{v}^T \exp[(\mathcal{L}_0 + \mathcal{L}_1)t] \mathbf{v} ,
\]
where the vector $\mathbf{v}$ is zero except on the eight terms of the form
\[
x = 000\alpha_1\alpha_2\alpha_3000\alpha_1\alpha_2\alpha_3 .
\]

\newpage

\subsection{The five-qubit code}
\label{app:five_qubit_code}

For the five-qubit code, we can write the dissipator in Eq.~(\ref{eq:Diss_five}) in matrix form in the basis of the $16$ density matrix components with different error weights given in Eq.~(\ref{eq:subdensity_5}):
\begin{widetext}
\begin{equation}\label{eq:L1_5q}
\mathcal{L}_1 = \gamma \mathcal{D}(\rho)=\gamma \begin{pmatrix}
-15& 1 & 0 & 0& 0& 0& 0& 0& 0& 0& 0& 0& 0& 0& 0& 0\\
 15&-13& 2 & 2& 0& 0& 0& 0& 0& 0& 0& 0& 0& 0& 0& 0\\
 0 & 4 &-15& 2& 3& 1& 0& 0& 0& 0& 0& 0& 0& 0& 0& 0\\
 0 & 8 & 4&-13& 0& 2& 3& 0& 0& 0& 0& 0& 0& 0& 0& 0\\
 0 & 0 & 3& 0&-15& 1& 0& 0& 1& 0& 4& 0& 0& 0& 0& 0\\
 0 & 0 & 6& 6 & 6&-12&6& 4& 3& 2& 0& 0& 0& 0& 0& 0\\
 0 & 0 & 0& 3 & 0& 2&-15& 0& 0& 2& 0& 0& 0& 0& 0& 0\\
 0 & 0 & 0& 0 & 0& 2& 0&-15& 3& 2& 0& 0& 3& 1& 0& 0\\
 0 & 0 & 0& 0 & 4& 2& 0& 4 &-14&2& 8& 4& 2& 0& 2& 0\\
 0 & 0 & 0& 0 & 0& 2& 6& 4 & 3&-11&0& 0& 0& 4& 0& 0\\ 
 0 & 0 & 0& 0 & 2& 0& 0& 0 & 1& 0&-15&1& 0& 0& 0& 5\\ 
 0 & 0 & 0& 0 & 0& 0& 0& 0 & 1& 0& 2&-14&2& 0& 2&10\\ 
 0 & 0 & 0& 0 & 0& 0& 0& 2 & 1& 0& 0& 4&-12&2&2&0\\ 
 0 & 0 & 0& 0 & 0& 0& 0& 1 & 0& 2& 0& 0& 3&-11& 6& 0\\ 
 0 & 0 & 0& 0 & 0& 0& 0& 0 & 1& 1& 0& 4& 2& 4& -15& 0\\ 
 0 & 0 & 0& 0 & 0& 0& 0& 0 & 0& 0& 1& 1& 0 &0 & 0& -15\\ 
\end{pmatrix} ,
\end{equation}
\end{widetext}
with eigenvalues $0$, $-4\gamma$, $-8\gamma$ (multiplicity $2$),$-12\gamma$ (multiplicity $3$),$-16\gamma$ (multiplicity $4$) and $-20\gamma$ (multiplicity $5$). The nullspace of the matrix is spanned by the vector
\[
(1,15,30,60,30,180,60,90,120,180,15,30,60,90,60,3)^T .
\]
The asymptotic state is the maximally mixed state, where all error states have the same probability. On the other hand, the error-correcting generator (with rate $\eta$) can be written
\begin{widetext}
\begin{equation}\label{eq:L0_5q}
[\mathcal{L}_0]=\eta \Gamma(\rho)=\eta \begin{pmatrix}
0& 1 & 0 & 0& 0& 0& 0& 0& 0& 0& 0& 0& 0& 0& 0& 0\\
0&-1 & 0 & 0& 0& 0& 0& 0& 0& 0& 0& 0& 0& 0& 0& 0\\
0& 0 &-1 & 0& 0& 0& 0& 0& 0& 0& 0& 0& 0& 0& 0& 0\\
0& 0 & 0 &-1& 0& 0& 0& 0& 0& 0& 0& 0& 0& 0& 0& 0\\
0& 0 & 0 & 0&-1& 0& 0& 0& 0& 0& 0& 0& 0& 0& 0& 0\\
0& 0 & 1 &1& 1&-1/3& 1& 2/3& 1/2& 1/3& 0& 0& 0& 0& 0& 0\\
0& 0 & 0 &0& 0& 0  &-1& 0& 0& 0& 0& 0& 0& 0& 0& 0\\
0& 0 & 0 &0& 0&1/3 & 0&-5/6&1/2&1/3& 0& 0& 1/2& 1/6& 0& 0\\
0& 0 & 0 &0& 0&0 & 0&0&-1&0& 0& 0& 0& 0& 0& 0\\
0& 0 & 0 &0& 0&0 & 0&0&0&-1& 0& 0& 0& 0& 0& 0\\
0& 0 & 0 &0& 0&0 & 0&0&0&0&-1&  0& 0& 0& 0& 0\\
0& 0 & 0 &0& 0&0 & 0&0&0&0&0&-1& 0& 0& 0& 0\\
0& 0 & 0 &0& 0&0 & 0&0&0&0&0&0&-1&  0& 0& 0\\
0& 0 & 0 &0& 0&0 & 0&1/6&0&1/3&0&0&1/2&-1/6&1& 0\\
0& 0 & 0 &0& 0&0 & 0&0&0&0&0&0&0&  0& -1& 0\\
0& 0 & 0 &0& 0&0 & 0&0&0&0&1&1&0& 0& 0& 0\\
\end{pmatrix} ,
\end{equation}
\end{widetext}
which has eigenvalues $0$, $-\eta$ and $(1/3)(-2\pm \sqrt{2})\eta$, and whose nullspace is spanned by the three vectors corresponding to $\rho_0$, $\rho_{5E}$, and
\[
(0, 0, 0, 0, 0, 2, 0, 1, 0, 0, 0, 0, 0, 1, 0, 0)^T ,
\]
associated with $2\rho_{3B}+\rho_{4A}+\rho_{5C}$.

\bibliography{GarciaBrun2023}

\end{document}